\documentclass[twocolumn,showpacs,prd]{revtex4}
\usepackage{mathrsfs}
\usepackage{longtable,lscape}
\usepackage{txfonts}
\usepackage{amssymb}
\usepackage{indentfirst}
\usepackage{graphicx,,booktabs}
\usepackage{color}
\usepackage{amssymb}

\begin{document}

\title{Categorizing resonances $X(1835)$, $X(2120)$ and $X(2370)$ in the pseudoscalar meson family}

\author{Jie-Sheng Yu$^{1,2}$}\author{Zhi-Feng Sun$^{1,2}$}
\author{Xiang Liu$^{1,2}$\footnote{Corresponding author}}\email{xiangliu@lzu.edu.cn}
\affiliation{$^1$School of Physical Science and Technology, Lanzhou University, Lanzhou 730000,  China\\
$^2$Research Center for Hadron and CSR Physics,
Lanzhou University and Institute of Modern Physics of CAS, Lanzhou 730000, China}

\author{Qiang Zhao$^{3,4}$}\email{zhaoq@ihep.ac.cn}
\affiliation{$3$Institute of High Energy Physics, Chinese Academy of Sciences, Beijing 100049, China\\
$^4$Theoretical Physics Center for Science Facilities, CAS, Beijing 100049, China}

\date{\today}

\begin{abstract}
Inspired by the newly observed three resonances $X(1835)$, $X(2120)$
and $X(2370)$, in this work we systematically study the two-body
strong decays and double pion decays of $\eta(1295)/\eta(1475)$,
$\eta(1760)/X(1835)$ and $X(2120)/X(2370)$ by categorizing
$\eta(1295)/\eta(1475)$, $\eta(1760)/X(1835)$, $X(2120)$ and
$X(2370)$ as the radial excitations of $\eta(548)/\eta^\prime(958)$.
Our numerical results indicate the followings: (1) The obtained
theoretical strong decay widths of three pseudoscalar states
$\eta(1295)$, $\eta(1475)$ and $\eta(1760)$ are consistent with the
experimental measurements; (2) $X(1835)$ could be the second radial
excitation of $\eta^\prime(958)$; (3) $X(2120)$ and $X(2370)$ can be
explained as the third and fourth radial excitations of
$\eta(548)/\eta^\prime(958)$, respectively.
The predicted two-body decay patterns of $\eta(1295)/\eta(1475)$,
$\eta(1760)/X(1835)$ and $X(2120)/X(2370)$ and their double pion
decays should be useful for further testing the conventional meson
assignment to $\eta(1295)/\eta(1475)$, $\eta(1760)/X(1835)$,
$X(2120)$ and $X(2370)$.
\end{abstract}

\pacs{14.40.Be, 13.25.Jx, 12.38.Lg} \maketitle

\section{Introduction}\label{sec1}

Very recently the BES-III Collaboration reported the observation of
several resonant structures in the $\eta^\prime \pi^+\pi^-$
invariant mass spectrum in $J/\psi\to \gamma \eta^\prime \pi^+\pi^-$
\cite{Collaboration:2010au}. Among these resonant structures,
$X(1835)$ that was first observed by the BES-II
Collaboration~\cite{Ablikim:2005um}, was confirmed with mass and
width of
$M_{X(1835)}=1836.5\pm3.0(\mathrm{stat})^{+5.6}_{-2.1}(\mathrm{syst})$
MeV and $\Gamma_{X(1835)}=190\pm
9(\mathrm{stat})^{+38}_{-36}(\mathrm{syst})$ MeV, respectively. The
BES-II determinations of the $X(1835)$ resonance parameters are
$M=1833.7\pm6.5(\mathrm{stat})\pm 2.7(\mathrm{syst})$ MeV and
$\Gamma=67.7\pm 20.3(\mathrm{stat})\pm7.7(\mathrm{syst})$ MeV with
$J^{PC}=0^{-+}$. In contrast with the BES-III result, it shows that
the newly measured width is larger than that from BES-II.

Apart from the $X(1835)$, the observation of another two resonant
structures initiate a lot of interests here, i.e.
\begin{eqnarray*}
M_{X(2120)}&=&2122.4\pm6.7(\mathrm{stat})^{+4.7}_{-2.7}(\mathrm{syst})\, \mathrm{MeV},\\\Gamma_{X(2120)}&=&83\pm16(\mathrm{stat})^{+31}_{-11}(\mathrm{syst})\, \mathrm{MeV},\\
M_{X(2370)}&=&2376.3\pm8.7(\mathrm{stat})^{+3.2}_{-4.3}(\mathrm{syst})\,
\mathrm{MeV},\\
\Gamma_{X(2370)}&=&83\pm17(\mathrm{stat})^{+44}_{-6}(\mathrm{syst})\,
\mathrm{MeV},
\end{eqnarray*}
where we refer to these two new structures by the names $X(2120)$
and $X(2370)$ in this work.

In 2005, when $X(1835)$ was firstly announced by the BES-II
Collaboration, it immediately stimulated tremendous interests in its
internal structure. In particular, its appearance in
$\eta^\prime\pi^+\pi^-$ instead of $\eta\pi^+\pi^-$ has initiated
various interpretations based on exotic configutations. Taking into
account the new observation of the $X(2120)$ and $X(2370)$ by
BES-III, it could be a great opportunity for us to gain some deep
insights into the isoscalar pseudoscalar meson spectrum. To proceed,
we first give a brief review of the studies of the $X(1835)$ in the
literature. We will then propose a classification scheme to
combining the $X(2120)$ and $X(2370)$ with the existing pseudoscalar
mesons in the quark pair creation (QPC) model.

In Ref.~\cite{Kochelev:2005vd}, $X(1835)$ was explained as the
lowest pseudoscalar glueball state due to the instanton mechanism of
partial $U(1)_A$ symmetry restoration. For further explaining why
the mass of $X(1835)$ is lower than the prediction of the quenched
Lattice QCD and QCD sum rules (QCDSR), an $\eta_c$-glueball mixing
mechanism was proposed \cite{Kochelev:2005tu}. Later, the authors of
Ref. \cite{He:2005nm} studied the QCD anomaly contribution to
$X(1835)$ using the QCDSR, and obtained a sizable matrix element
$\langle 0|G\tilde{G}|G_P \rangle$ for $X(1835)$. It shows that
$X(1835)$ could be explained as a pseudoscalar state with a large
gluon content. In Ref. \cite{Li:2005vd}, a further study of
$X(1835)$ as a pseudoscalar glueball was performed by estimating the
decay rates of $X(1835)\to VV,\, \gamma V,\,\gamma\gamma$ and cross
sections of $\gamma\gamma\to X(1835)\to f$ and $h_1+h_2\to
X(1835)+\cdots$ by an effective Lagrangian approach. A $0^{-+}$
trigluon glueball with mass range $1.9\sim 2.7$ GeV was also
investigated within a baryonium-gluonium mixing picture using the
QCDSR \cite{Hao:2005hu}.

\begin{figure*}[hbtp]
\begin{center}
\begin{tabular}{c}
\includegraphics[height=200pt]{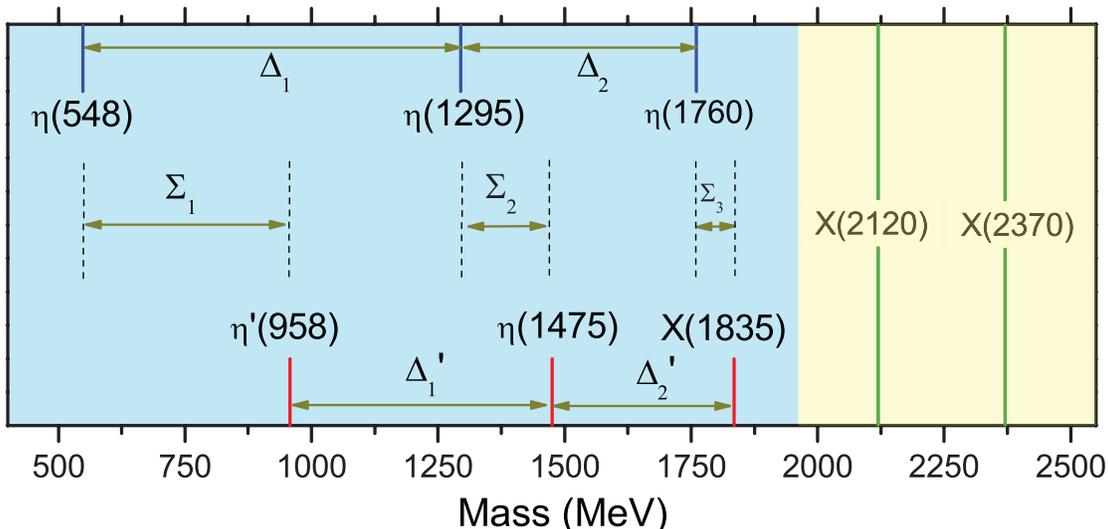}\\ [-20pt]
\end{tabular}
\end{center}
\caption{(Color online.)  A summary of the isoscalar pseudoscalar
states assuming $X(2120)$ and $X(2370)$ as pseudoscalar states.
Here, all data are taken from PDG \cite{Amsler:2008zzb} and the BES
observations \cite{Ablikim:2005um,Collaboration:2010au}.
\label{review}}
\end{figure*}
\refstepcounter{figure}

It is worth mentioning that before the observation of $X(1835)$, the
BES-II Collaboration once reported a  $p\bar{p}$ subthreshold
enhancement $X(1860)$ in the $J/\psi\to \gamma p\bar{p}$ decay
\cite{Bai:2003sw}. This observation has also stimulated broad
studies of the nature of this enhancement among which the $p\bar{p}$
baryonium appears to be attractive
\cite{Datta:2003iy,Gao:2003ka,Yan:2004xs,Liu:2004er}. The new data
from the BES-III Collaboration also confirmed the $X(1860)$ signal
in $\psi^\prime\to\pi^+\pi^-J/\psi(J/\psi\to\gamma p\bar{p})$
\cite{Collaboration:2010zd}. Later, the CLEO Collaboration also
reported the observation of $X(1860)$ \cite{Alexander:2010vd} as an
independent confirmation of this enhancement. Combining the
$X(1860)$ signal with the later observed $X(1835)$ by BES-II, it was
conjectured that these two structures might be originated from the
same resonant state.
Ding and Yan proposed that $X(1835)$ could be treated as a baryonium
with a sizable gluon content. This would explain why $X(1835)$ was
produced in the $J/\psi$ radiative decays and with large couplings
to $p\bar p,\, \eta^\prime \pi\pi$
\cite{Ding:2005gh,Yan:2006ht,Ding:2007xi}. Calculations by the QCDSR
\cite{Wang:2006sna} and large-$N_c$ QCD \cite{Liu:2007tj} seemed to
support that $X(1835)$ contained a baryonium component. A study
based on the conventional $N\bar{N}$ potential model
\cite{Dedonder:2009bk} suggested that $X(1835)$ could be a broad and
weakly bound state $N\bar{N}_S(1870)$ in the $^{11}S_0$ wave.
However, the calculations of the strong and electromagnetic decays
of $X(1835)$ under the baryonium assumption obtained a rather small
width for $X(1835)\to \eta^\prime \pi^+\pi^-$ \cite{Ma:2008hc}.

Apart from those proposed explanations for $X(1835)$ as an exotic
state, efforts have also been made to understand $X(1835)$ from the
point of view of a conventional $q\bar q$ state. Huang and Zhu
treated $X(1835)$ as the second radial excitation of
$\eta^\prime(958)$ and discussed the strong decay behavior by the
effective Lagrangian approach \cite{Huang:2005bc}. In Ref.
\cite{Li:2008mza}, several two-body strong decays of $X(1835)$
associated with $\eta(1760)$ were studied by the QPC model, where
$X(1835)$ is assigned as the $n^{2s+1}L_J=3^1S_0$ state.

Although great efforts have been made in the literature, the
properties of $X(1835)$ still remain unclear at present. This
situation may be improved by the BES-III observations of $X(2120)$
and $X(2370)$ associated with $X(1835)$ in the $\eta^\prime
\pi^+\pi^-$ invariant mass spectrum. Nevertheless, the upgraded
experimental information may allow us to have an overall view of the
pseudoscalar meson spectrum, which has not been systematically
addressed before.

To proceed, we organize the paper as follows. After the
Introduction, we propose a quark model scheme to organize the so-far
observed pseudoscalar mesons based on a qualitative analysis of the
pseudoscalar mass spectrum. In Sec. \ref{sec3}, the QPC model for
the study of the strong decays of $\eta(1295)$, $\eta(1475)$,
$\eta(1760)$, $X(1835)$, $X(2120)$ and $X(2370)$ in our categorizing
scheme is summarized. In Sec. \ref{sec4}, the numerical results are
presented and compared with the experimental data. Discussion and
conclusion are given in Sec. \ref{sec5}.

\section{The categorizing scheme for the pseudoscalar meson spectrum}\label{sec2}

In particle data group (PDG)~\cite{Amsler:2008zzb}, six isoscalar
pseudoscalar states, $\eta(548)$, $\eta^\prime(958)$, $\eta(1295)$,
$\eta(1405)$, $\eta(1475)$, $\eta(1760)$ and $\eta(2225)$, are
listed as observed states. Among these states, $\eta(548)$ and
$\eta^\prime(958)$ are well established as the ground states of the
$0^{-}$ nonet associated with $\pi^0$, $\pi^\pm$, $K^\pm$, $K^0$ and
$\bar{K^0}$. Such a quark model scenario should be a good starting
point for classifying the observed higher pseudoscalar states as
radial excitations of the ground states.

It is well-established that $\pi^{\pm,0}$, $K^{\pm,0}$, $\bar{K}^0$,
$\eta(548)$ and $\eta^\prime(958)$ belong to the ground state
pseudoscalar nonet. It is also recognizable that $\pi(1300)$,
$K(1460)$, $\eta(1295)$, and $\eta(1475)$ make the first radial
excitation states of $0^-$ mesons~\cite{Klempt:2007cp}. There have
been broad discussions about the nature of $\eta(1405)$. For
instance, a recent review of this state can be found in
Refs.~\cite{Masoni:2006rz,Amsler:2008zzb,Bugg:2011tr}.
In
most theoretical studies \cite{Cheng:2008ss,Gerasimov:2007sb,Gutsche:2009jh,Close:1996yc,Li:2003cn,Li:2007ky},
the decay patterns of $\eta(1405)$ indicate its being possible
candidate of $0^{-+}$ glueball with the mass consistent with the
prediction of the flux tube model \cite{Faddeev:2003aw}. In any
case, if one leaves $\eta(1405)$ as a pseudoscalar glueball and to
be investigated separately, it is interesting to recognize a
pseudoscalar nonet of the second radial excitations formed by
$\pi(1800)$, $K(1830)$, $\eta(1760)$ and $X(1835)$. Note that
$\eta(1760)$ was observed in the invariant mass spectra of
$\omega\omega$~\cite{Ablikim:2006ca} and
$\rho\rho$~\cite{Bisello:1988as}. It makes a natural assignment of
$\eta(1760)$ as the second radial excitation of $\eta(548)$. In
contrast, $X(1835)$ strongly couples to $\eta^\prime\pi^+\pi^-$
instead of $\eta\pi^+\pi^-$. This makes it a reasonable partner of
$\eta(1760)$ as the second radial excitation of $\eta^\prime(958)$
\cite{Huang:2005bc}.

\begin{table}[htb]
\centering%
\caption{The pseudoscalar nonet. \label{nonet}}
\begin{tabular}{cccccccc}
\toprule[1pt]
$1S$&$2S$&$3S$&$4S$&$5S$\\\midrule[1pt]
$\eta,\,\eta^\prime$&$\eta(1295)$&$\eta(1760)$& $X(2120)$&$X(2370)$\\
                    &$\eta(1475)$&$X(1835)$&&    \\
$K(494)$&$K(1460)$&$K(1830)$&&\\
$\pi$ &$\pi(1300)$&$\pi(1800)$&&\\
\bottomrule[1pt]
\end{tabular}
\end{table}

Since $X(2120)$ and $X(2370)$ associated with $X(1835)$ were
observed in the $\eta^\prime \pi\pi$ mass spectrum, we naturally
deduce that the newly observed $X(2120)$ and $X(2370)$ could be
categorized as the radial excitation states in $\eta-\eta^\prime$
family. There exist several possible assignments to $X(2120)$ and
$X(2370)$ due to lack of further experimental information: (1)
$X(2120)$ is the third radial excitation of $\eta(548)$ or
$\eta^\prime(958)$; (2) $X(2370)$ is the fourth radial excitation of
$\eta(548)$ or $\eta^\prime(958)$; (3) $X(2120)$ and $X(2370)$ are
the third radial excitations of $\eta(548)$ and $\eta^\prime(958)$,
respectively.

The above categorizing can also be understood by examining the mass
spectrum. In Fig. \ref{review}, we present the mass spectrum of all
these states with their mass gaps indicated explicitly by
$\Delta_{1}^{(\prime)}$, $\Delta_2^{(\prime)}$, and $\Sigma_{i}$
with $(i=1,2,3)$. The qualitative relations, $\Delta_2<\Delta_1$,
$\Delta_2^\prime<\Delta_1^\prime$ and $\Sigma_3<\Sigma_2<\Sigma_1$,
are consistent with the expectations of constituent quark model. It
supports the assignment to $X(1835)$ as the second radial excitation
of $\eta^\prime(958)$.

Following the above relations, we can apply $\Delta_{j}<\Delta_i$
and $\Delta_{j}^\prime<\Delta_i^\prime$ with ($j>i$) as a criteria
to distinguish the different assignments to $X(2120)$ and $X(2370)$.
$X(2120)$ as the third radial excitation of $\eta(548)$ or
$\eta^\prime(958)$ can result in the mass gap between $X(2120)$ and
$\eta(1760)(X(1835))$ is smaller than the corresponding
$\Delta_2(\Delta_2^\prime)$. $X(2370)$ as the fourth radial
excitation of $\eta(548)$ is suitable since the mass gap between
$X(2370)$ and $X(2120)$ is smaller than that between $X(2120)$ and
$\eta(1760)$. Additionally, $X(2370)$ could be as the fourth radial
excitation of $\eta^{\prime}(958)$, which will be discussed later.

\begin{center}
\begin{figure}[htb]
\begin{tabular}{c}
\scalebox{0.97}{\includegraphics{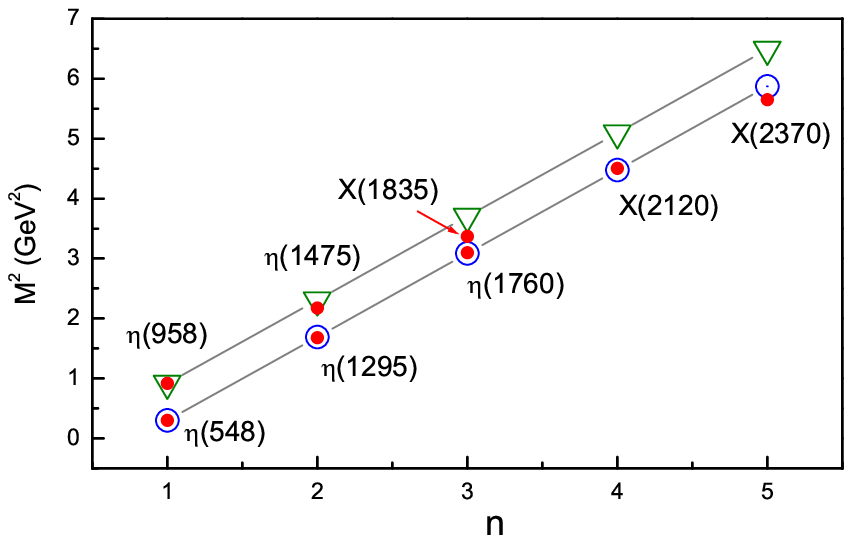}}
\end{tabular}
\caption{The Regge trajectories for the $\eta/\eta^\prime$ mass
spectrum with $M^2=M_0^2+(n-1)\mu^2$ ($\mu^2=1.39$ GeV)
\cite{Anisovich:2000kxa}. The $\eta$ and $\eta^\prime$ trajectories
are marked by "\textcolor[rgb]{0.00,0.00,1.00}{$\odot$}" and
"\textcolor[rgb]{0.00,0.50,0.00}{$\bigtriangledown$}", respectively.
The red points are experimental data from the PDG
\cite{Amsler:2008zzb}. \label{RT}}
\end{figure}
\end{center}

The above assignments of $X(1835)$,
$X(2120)$ and $X(2370)$ seem to be consistent with the analysis of
Regge trajectories (RTs) \cite{Chew:1962eu}. This is an valuable
approach to provide quantitative estimate of hadron masses with the
same quantum number. In Fig. \ref{RT}, we plot the $0^{-+}$
trajectory on the plane of $(n, \ M^2)$ adopting the relation
$M^2=M_0^2+(n-1)\mu^2$ from Ref. \cite{Anisovich:2000kxa}, where
$M_0$ is the ground state mass, $n$ the radial quantum number, and
$\mu^2$ the slope parameter of the trajectory. It shows that
$X(1835)$, $X(2120)$ and $X(2370)$ can be well accommodated into the
trajectory.

Our assignment  to $X(2370)$ and $X(2120)$ here is different from
that suggested in Ref. \cite{Liu:2010tr}, where $X(2120)$ and
$X(2370)$ are assumed as the third radial excitations of $\eta(548)$
and $\eta^\prime(958)$, respectively. In  Ref. \cite{Liu:2010tr}, it
was concluded that $X(2120)$ and $X(2370)$ can not be understood as
the third radial excitations of $\eta(548)$ and $\eta'(958)$ while
$X(2370)$ is probably a mixture of $\eta'(4{^1}{S}{_0})$ and
glueball. In this work, we expect that a systematic study of the
two-body and double pion strong decays of all these states would
provide useful information for our understanding of the $\eta$ and
$\eta'$ spectrum.

\section{Two-body and three-body strong decays}\label{sec3}

\subsection{Quark pair creation model}

In the following, we give a brief review of the QPC model (also
known as $^3P_0$ model) adopted in this work for the study of the
strong decays of the states in the $\eta-\eta^\prime$ family. Early
developments of this method can be found in the literature
\cite{Micu:1968mk,yaouanc,LeYaouanc:1977gm,LeYaouanc:1988fx,vanBeveren:1979bd,Bonnaz:2001aj,sb}.
It has also been broadly applied to the study of hadron properties
related to recent progresses on the hadron spectroscopy
\cite{Li:2008mza,Blundell:1995ev,Page:1995rh,Capstick:1986bm,Capstick:1993kb,Ackleh:1996yt,Zhou:2004mw,Guo:2005cs,
Close:2005se,Lu:2006ry,Zhang:2006yj,Chen:2007xf,Sun:2009tg,Liu:2009fe,Sun:2010pg,Li:2008mzb,Luo:2009wu}.

\begin{center}
\begin{figure}[htb]
\begin{tabular}{c}
\scalebox{1.3}{\includegraphics{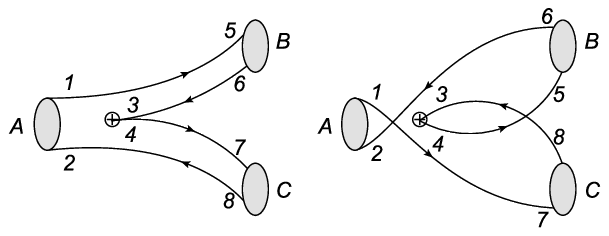}}
\end{tabular}
\caption{The quark level diagrams describing meson decay in QPC model. \label{qpc}}
\end{figure}
\end{center}

In the QPC model, a transition operator $T$  is introduced to
describe a quark-antiquark pair creation from the vacuum
\begin{eqnarray}
T&=& - 3 \gamma \sum_m\: \langle 1\;m;1\;-m|0\;0 \rangle\,
\int\!{\rm d}{\textbf{k}}_3\; {\rm
d}{\textbf{k}}_4 \delta^3({\textbf{k}}_3+{\textbf{k}}_4) \nonumber\\&&\times{\cal
Y}_{1m}\left(\frac{{\textbf{k}}_3-{\textbf{k}_4}}{2}\right)\;
\chi^{3 4}_{1, -\!m}\; \varphi^{3 4}_0\;\,
\omega^{3 4}_0\; d^\dagger_{3i}({\textbf{k}}_3)\;
b^\dagger_{4j}({\textbf{k}}_4)\,,\label{tmatrix}
\end{eqnarray}
where dimensionless parameter $\gamma$ denotes  the creation
strength of a quark-antiquark pair with quantum number
$J^{PC}=0^{++}$. $i$ and $j$ are the $SU(3)$ color indices of the
created quark and anti-quark. $\varphi^{34}_{0}=(u\bar u +d\bar d +s
\bar s)/\sqrt 3$ and
$\omega_{0}^{34}=\frac{1}{\sqrt{3}}\delta_{\alpha_3\alpha_4}\,(\alpha=1,2,3)$
correspond to flavor and color singlets, respectively.
$\chi_{{1,-m}}^{34}$ denotes a triplet state of spin.
$\mathcal{Y}_{\ell m}(\mathbf{k})\equiv |\mathbf{k}|^{\ell}Y_{\ell
m}(\theta_{k},\phi_{k})$ is the $\ell$th solid harmonic polynomial.
The schematic diagrams in Fig. \ref{qpc} illustrate the decay
transitions via the quark pair creation process. Note that  the
right diagram in Fig. \ref{qpc} is valid only when the quark
components in mesons $A$, $B$ and $C$ are the same as each other.

The expression of decay width in the QPC model is written as
\begin{eqnarray}
\Gamma = \pi^2 \frac{{|\textbf{K}|}}{M_A^2}\sum_{JL}\Big
|\mathcal{M}^{J L}\Big|^2,
\end{eqnarray}
where $\mathcal{M}^{J L}$ is the partial wave amplitude and related
to the helicity amplitude $\mathcal{M}^{M_{J_A} M_{J_B} M_{J_C}}$
according to the Jacob-Wick  formula \cite{Jacob:1959at}. The
helicity amplitude $\mathcal{M}^{M_{J_A} M_{J_B} M_{J_C}}$ is
obtained by the transition amplitude
\begin{eqnarray}
\langle BC|T|A\rangle=\delta^3(\mathbf{K}_B+\mathbf{K}_C-\mathbf{K}_A)\mathcal{M}^{M_{J_A}M_{J_B}M_{J_C}}.
\end{eqnarray}
A detailed review of the QPC  model and calculation of the
transition amplitude $\langle BC|T|A\rangle$ have been given in Ref.
\cite{Luo:2009wu}, where the simple harmonic oscillator (SHO) wave
function is applied to describe the meson spatial wave function
\begin{eqnarray}
\Psi_{nLM}(\mathbf{k})=\mathcal{N}_{nL}
\exp\left(-\frac{R^2\mathbf{k}^2}{2}\right)\mathcal{Y}_{LM}(\mathbf{k})\,\mathcal{P}(\mathbf{k}^2),
\label{sho}
\end{eqnarray}
where $\mathcal{P}(\mathbf{k}^2)$ is a polynomial in terms of
$\mathbf{k}^2$, the relative momentum between the quark and the
anti-quark within a meson, and $\mathcal{N}_{nL}$ represents the
normalization coefficient.

\begin{figure*}[hbtp]
\begin{center}
\begin{tabular}{c}
\includegraphics[height=600pt]{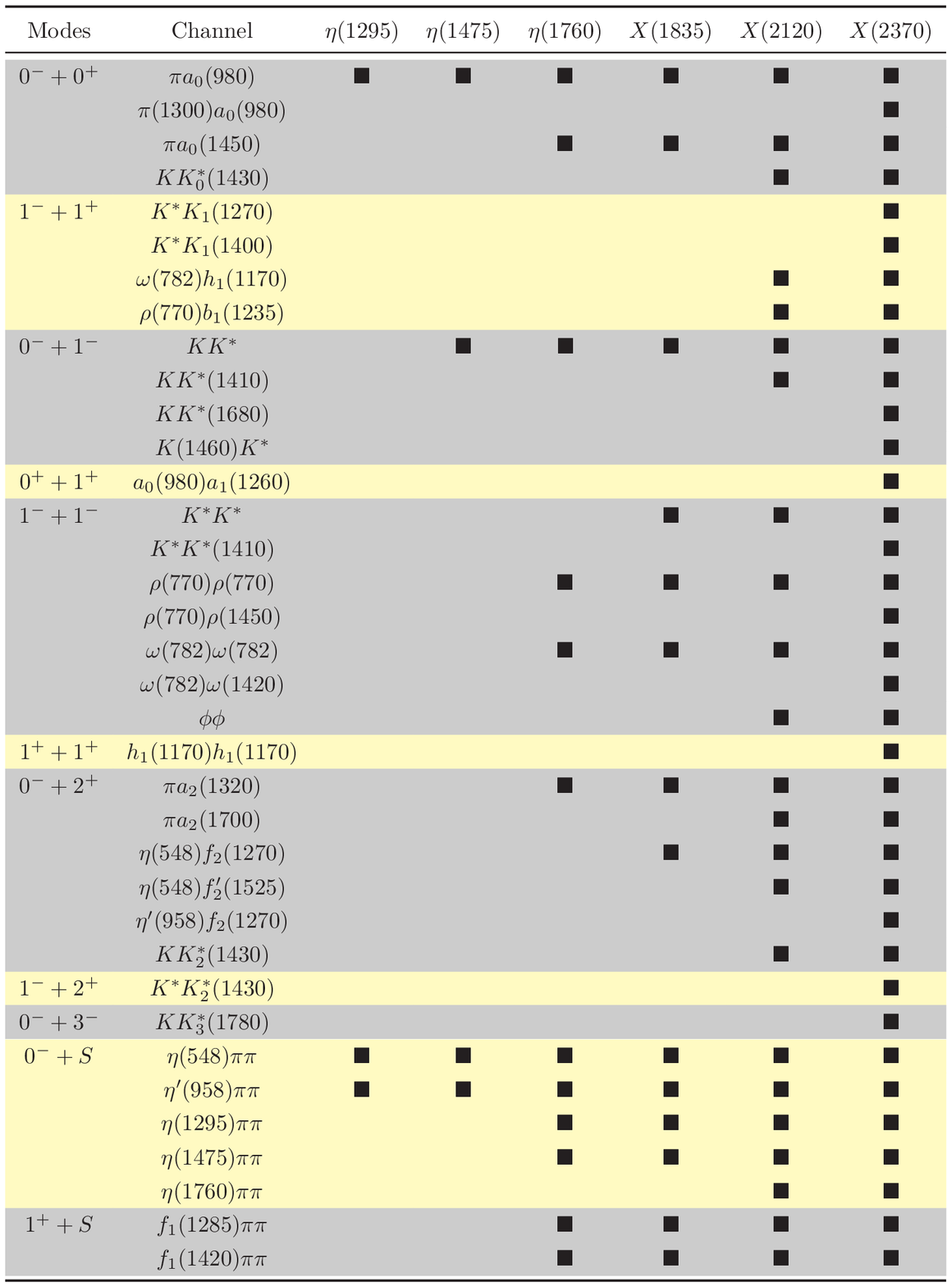}\\ [-20pt]
\end{tabular}
\end{center}
\caption{The two-body decays and double pion decays of
$\eta(1295/\eta(1475)$, $\eta(1760)/X(1835)$, $X(2120)/X(2370)$. The
double pion decays occur via the intermediate scalar mesons ($S$),
such as $\sigma(600)$ and $f_0(980)$. Here, we use $\blacksquare$ to
mark the allowed two-body decays and double pion decays of
$\eta(1295/\eta(1475)$, $\eta(1760)/X(1835)$, $X(2120)/X(2370)$
\label{decaychannel}}
\end{figure*}
\refstepcounter{figure}

\renewcommand{\arraystretch}{1.6}
\begin{table*}[hbtp]
\begin{tabular}{c|c}\toprule[1pt]
Decay modes&Partial wave amplitude\\\midrule[1pt]

$0^{-}\to0^-+0^+$&$\mathcal{M}^{00}=\frac{\sqrt{2}}{3}\mathcal{F}\sqrt{E_A E_B E_C}\gamma[\mathcal{I}^{0-1}_{0-1}(nS)+\mathcal{I}^{00}_{00}(nS)+\mathcal{I}^{01}_{01}(nS)]$\\
$0^{-}\to1^-+1^+(^{1}P_1)$&$\mathcal{M}^{00}=\frac{\sqrt{2}}{3}\mathcal{F}\sqrt{E_A E_B E_C}\gamma[\mathcal{I}^{0-1}_{0-1}(nS)+\mathcal{I}^{00}_{00}(nS)+\mathcal{I}^{01}_{01}(nS)]$\\
$0^{-}\to1^-+1^+(^{3}P_1)$&$\mathcal{M}^{00}=\frac{2}{3}\mathcal{F}\sqrt{E_A E_B E_C}\gamma[\mathcal{I}^{0-1}_{0-1}(nS)+\mathcal{I}^{00}_{00}(nS)+\mathcal{I}^{01}_{01}(nS)]$\\\midrule[1pt]

$0^{-}\to0^-+1^-$&$\mathcal{M}^{11}=\sqrt{\frac{2}{3}}\mathcal{F}\sqrt{E_A E_B E_C}\gamma{\mathcal{Q}^{00}_{00}(nS)}$\\
$0^{-}\to0^++1^+(^{1}P_1)$&$\mathcal{M}^{11}=-\frac{\sqrt{2}}{3}\mathcal{F}\sqrt{E_A E_B E_C}\gamma[\mathcal{U}^{0-1}_{-10}(nS)+\mathcal{U}^{00}_{00}(nS)+\mathcal{U}^{01}_{10}(nS)]$\\
$0^{-}\to0^++1^+(^{3}P_1)$&$\mathcal{M}^{11}=-\mathcal{F}\frac{1}{3}\sqrt{E_A E_B E_C}\gamma[\mathcal{U}^{0-1}_{0-1}(nS)+\mathcal{U}^{00}_{-11}(nS)+\mathcal{U}^{00}_{1-1}(nS)+\mathcal{U}^{01}_{01}(nS)]$\\
$0^{-}\to1^+(^{1}P_1)+1^+(^{1}P_1)$&$^\star\mathcal{M}^{11}=0$\\
$0^{-}\to1^-+1^-$&$\mathcal{M}^{11}=-\frac{2}{\sqrt{3}}\mathcal{F}\sqrt{E_AE_BE_C}\gamma{\mathcal{Q}^{00}_{00}(nS)}$\\\midrule[1pt]

$0^{-}\to0^-+2^+$&$\mathcal{M}^{22}={\frac{1}{3}}\mathcal{F}\sqrt{E_A E_B E_C}\gamma[\mathcal{I}^{0-1}_{0-1}(nS)-2\mathcal{I}^{00}_{00}(nS)+\mathcal{I}^{01}_{01}(nS)]$\\
$0^{-}\to0^++2^-(^{1}D_2)$&$\mathcal{M}^{22}=\frac{\sqrt{2}}{3}\mathcal{F}\sqrt{E_A E_B E_C}\gamma[\mathcal{G}^{0-1}_{-10}(nS)+\mathcal{G}^{00}_{00}(nS)+\mathcal{G}^{01}_{10}(nS)]$\\
$0^{-}\to0^++2^-(^{3}D_2)$&$\mathcal{M}^{22}=\frac{1}{3}\mathcal{F}\sqrt{E_A E_B E_C}\gamma[\mathcal{G}^{0-1}_{0-1}(nS)+\mathcal{G}^{00}_{-11}(nS)+\mathcal{G}^{00}_{1-1}(nS)+\mathcal{G}^{01}_{01}(nS)]$\\
$0^{-}\to1^-+2^+$&$\mathcal{M}^{22}=-\frac{1}{\sqrt{6}}\mathcal{F}\sqrt{E_A E_B E_C}\gamma[\mathcal{I}^{0-1}_{0-1}(nS)-2\mathcal{I}^{00}_{00}(nS)+\mathcal{I}^{01}_{01}(nS)]$\\
$0^{-}\to1^-+1^+(^{1}P_1)$&$\mathcal{M}^{22}={\frac{1}{3}}\mathcal{F}\sqrt{E_A E_B E_C}\gamma[\mathcal{I}^{0-1}_{0-1}(nS)-2\mathcal{I}^{00}_{00}(nS)+\mathcal{I}^{01}_{01}(nS)]$\\
$0^{-}\to1^-+1^+(^{3}P_1)$&$\mathcal{M}^{22}=-{\frac{1}{3\sqrt{2}}}\mathcal{F}\sqrt{E_A E_B E_C}\gamma[\mathcal{I}^{0-1}_{0-1}(nS)-2\mathcal{I}^{00}_{00}(nS)+\mathcal{I}^{01}_{01}(nS)]$\\
\bottomrule[1pt]
\end{tabular}\caption{The general expressions of the partial wave amplitudes for the strong decays of
$\eta(1295)/\eta(1475)$, $\eta(1760)/X(1835)$ and $X(2120)/X(2380)$.
Here, $\mathcal{F}$ is the relevant flavor matrix element. Due to
the complication of the concrete expressions of
$\mathcal{I}_{ij}^{k\ell}(nS)$ and $\mathcal{Q}_{ij}^{k\ell}(nS)$,
the detailed formulae of $\mathcal{I}_{ij}^{k\ell}(nS)$ and
$\mathcal{Q}_{ij}^{k\ell}(nS)$ are not given in this work. $E_\beta$
($\beta=A,\,B,\,C$) are the energies of the initial and final
states. \label{amp}}
\end{table*}

\subsection{Strong decay behavior}

Before presenting the calculation results for the strong decays of
$\eta(1295)$, $\eta(1440)$, $\eta(1760)$, $X(1835)$, $X(2120)$,
$X(2370)$, we briefly introduce the mixing scheme of the
$\eta-\eta^\prime$ family. In the SU(3) quark model, the physical
states $\eta(nS)$ and $\eta^\prime(nS)$ with the same radial
excitation quantum number could be the mixtures of $\eta_{q}(nS)$
and $\eta_{s}(nS)$ in the flavor basis
\begin{eqnarray}\label{mixing1}
 \left ( \begin{array}{ccc} \eta(nS)\\  \eta^\prime(nS)\end{array} \right )=
 \left ( \begin{array}{ccc}
  \rm \cos\theta_n & \rm -\sin\theta_n \\
  \rm \sin\theta_n & \rm \cos\theta_n \end{array} \right )
 \left ( \begin{array}{ccc} \eta_{q}(nS)  \\  \eta_{s}(nS)  \end{array}\right
),
\end{eqnarray}
where $\eta_{q}(nS)$ and $\eta_{s}(nS)$ are the flavor wave
functions $|\eta_{q}(nS)\rangle ={1\over\sqrt 2}(|u\bar
u\rangle+|d\bar d\rangle)$ and $|\eta_{s}(nS)\rangle =|s\bar
s\rangle$, respectively. For instance, $\eta(548)/\eta^\prime(958)$
are the ground states with $n=1$, for which the commonly adopted
mixing angle is $\theta_1=(39.3\pm1.0)^\circ$
\cite{Feldmann:1998vh}. For the first radial excitations,
$\eta(1295)$ and $\eta(1475)$ are organized as the physical states
with mixing angle $\theta_2$. For $\eta(1760)/X(1835)$ and
$X(2120)/X(2370)$ to be discussed in the following subsections,
mixing angles $\theta_3$, $\theta_4$ and $\theta_5$ are introduced
respectively. In contrast with the better determined mixing angle
$\theta_1$, information about other mixing angles $\theta_\alpha $
($\alpha=2,3,4,5$) is still absent. We expect that the
$\theta_\alpha$-dependence of the resonance decay widths may provide
some constraints on the mixing angles.

The two-body and double pion decay channels which are allowed by the
conservation law are listed in Table \ref{decaychannel} for
$\eta(1295)$, $\eta(1440)$, $\eta(1760)$, $X(1835)$, $X(2120)$, and
$X(2370)$. We assume that the double pion decay occurs through the
intermediate scalar mesons, such as $\sigma$ and $f_0(980)$.

By the QPC model, the general expressions of the partial wave
amplitudes for the strong decays of $\eta(1295)/\eta(1440)$,
$\eta(1760)/X(1835)$, $X(2120)/X(2370)$ are obtained and listed in
Table \ref{amp}. $\Xi_{ij}^{k\ell}(nS)$ $(\Xi=\mathcal{I,U,Q,G})$ is
extracted from the spatial integral, which describes the overlap of
the initial pseudoscalar meson with the radial quantum number $n$
and the created pair within the two final mesons. Here, the simple
harmonic oscillator (SHO) wave function $\Psi_{n \ell
m}(\mathbf{k})=R_{n\ell m}(R,\mathbf{k})\mathcal{Y}_{n\ell
m}(\mathbf{k})$ is introduced in the calculation of the spatial
integrals.

The input parameters,
which include the $R$ values in the SHO wave functions, flavor wave
functions, masses, and the creation strength of a quark-antiquark
pair from vacuum, are collected in Table. \ref{parameter}.

For estimating the double pion decay widths, we assume that the
double pion decays listed in Fig. \ref{decaychannel} occur through
the intermediate scalar mesons $\sigma$ and $f_0(980)$. The general
expression of the double pion decay is \cite{Luo:2009wu}
\begin{eqnarray}
&&\Gamma(X\to \eta+\mathcal{S}\to\eta+\pi\pi)\nonumber\\&&=\sum_{\mathcal{S}=\sigma,f_0}\frac{1}{\pi}
\int_{4m_\pi^2}^{(m_X-m_\eta)^2}dr\sqrt{r}\frac{\Gamma_{X\to \eta+\mathcal{S}}(r)\cdot \Gamma_{\mathcal{S}\to \pi\pi}(r)}{(r-{m_{\mathcal{S}}^2})^2+(m_{\mathcal{S}}\Gamma_{\mathcal{S}})^2}\label{11},
\end{eqnarray}
where $X$ denotes any of states in the $\eta-\eta^\prime$ family.
$\Gamma_{X\to \eta+\mathcal{S}}(r)$ is
obtained in the QPC model. The effective Lagrangian describing the
interaction of scalar state ($\mathcal{S}=\sigma,\,f_0(980)$) with
two pions is written as
\begin{eqnarray}
\mathcal{L}_{\mathcal{S}\pi\pi}&=&g_{\sigma}\sigma(2\pi^+\pi^-+\pi^0\pi^0),
\end{eqnarray}
where the coupling constants $g_\sigma=2.60\sim3.35$ GeV and
$g_{f_0}=0.83\sim1.30$ GeV  are determined by the total widths of
$\sigma$ and $f_{0}(980)$, i.e. $\Gamma_\sigma=600\sim1000$ MeV and
$\Gamma_{f_0}=40\sim 100$ MeV \cite{Amsler:2008zzb}. The decay
amplitudes of $\mathcal{S}\to \pi\pi$ are
\begin{eqnarray}
\Gamma_{\sigma\to\pi\pi}=\frac{g_{\sigma}^2\lambda^2}{8\pi}\frac{p_1(r)}{r}
\end{eqnarray}
with $p_1(r)$ denotes the three-vector momentum of the final state
pion in the initial scalar rest frame.


In this work, we also discuss the sequential decay $X\to
\pi+a_0(980)\to \eta\pi \pi$. The general expression of the decay
width of $X\to \pi^0+a_0(980)^0\to \eta\pi^0 \pi^0$ is similar to
Eq. (11). It reads
\begin{eqnarray}
&&\Gamma(X\to \pi^0+a_0(980)^0\to
\eta\pi^0\pi^0)\nonumber\\&&\approx\frac{1}{\pi}
\int_{(m_{\pi^0}+m_\eta)^2}^{(m_X-m_{a^0})^2}dr\sqrt{r}
\frac{\Gamma_{X\to \pi^0+a_0^0}(r)\cdot \Gamma_{a_0^0\to
\pi^0\pi^0}(r)}{(r-{m_{a_0}^2})^2+(m_{a_0}\Gamma_{a_0})^2},\label{111}
\end{eqnarray}
where the decay width $\Gamma_{a_0^0\to
\eta\pi^0}=({g_{a_0}^2}/{(8\pi r)})
((r-(m_\pi+m_\eta)(r-(m_\pi-m_\eta))/(2r^{1/2})$ with
$g_{a_0}=1.262\sim 2.524$ GeV determined by the total decay width of
$a_0(980)$ ($\Gamma_{a_0}=50\sim 100$ MeV). The final result of
$\Gamma_{X\to \pi+a_0(980)\to \eta\pi \pi}$ includes the
contributions from both $\eta\pi^+\pi^-$  and $\eta\pi^0\pi^0$.

\section{numerical result}\label{sec4}

In this Section, the numerical results for states $\eta(1295)$,
$\eta(1475)$, $\eta(1760)$, $X(1835)$, $X(2120)$ and $X(2370)$, are
presented by pairing them as the $\eta(nS)/\eta^\prime(nS)$
partners. In the QPC model, the predicted partial widths will have
dependence on the harmonic oscillator strength $R$. Thus, we will
illustrate the partial widths in terms of $R$. By adopting the data
for some of those measured channels, we can then examine the model
predictions within a fixed range of $R$.

\subsubsection{$\eta(1295)$ and $\eta(1475)$}

\begin{center}
\begin{figure*}[htbp]
\begin{tabular}{cc}
\scalebox{0.96}{\includegraphics{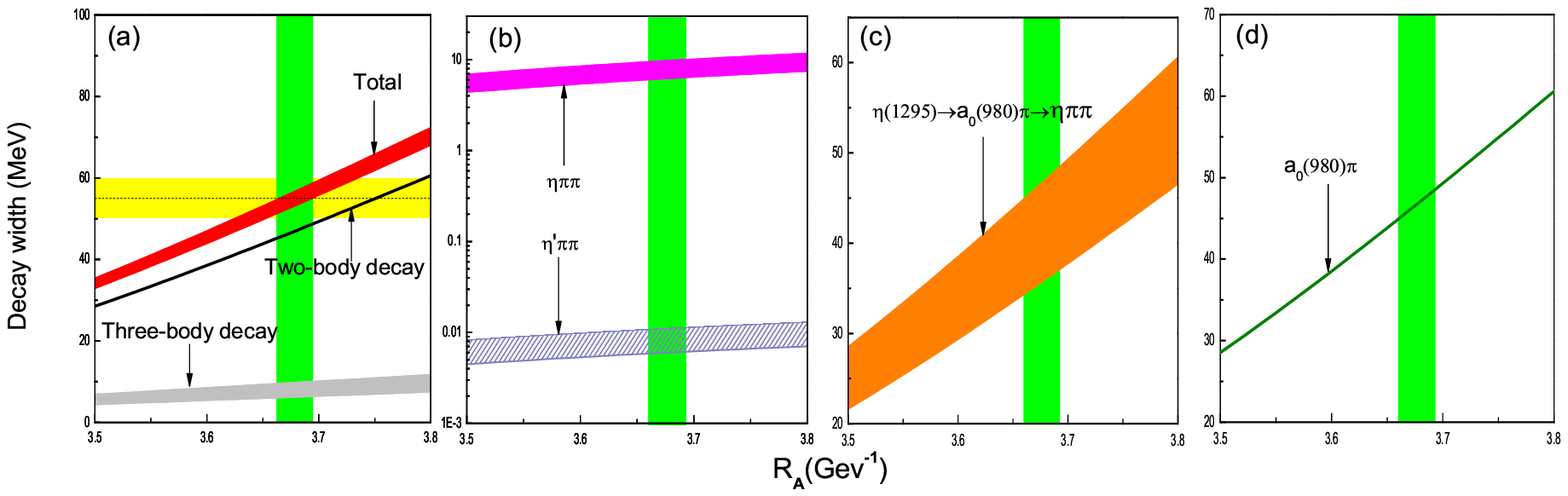}}
\end{tabular}
\caption{ (Color online). The $R$ dependence of the calculated total
and partial widths of $\eta(1295)$.  (a) The total decay width,
two-body and three-body strong decay of $\eta(1295)$ in comparison
with the experimental data (dashed line with yellow band). (b) The
$R$ dependence of the double pion decays of $\eta(1295)$ via the
intermediate scalar states. (c) The decay width of $\eta(1295)\to
a_0(980)\pi\to \eta\pi\pi$. (d) The partial two-body decay width of
$\eta(1295)$. Here, the $R$ range corresponding to the overlap
between the total decay width and experimental data is marked by the
green band.}\label{1295}
\end{figure*}
\end{center}

\begin{center}
\begin{figure*}[htbp]
\begin{tabular}{cc}
\scalebox{0.89}{\includegraphics{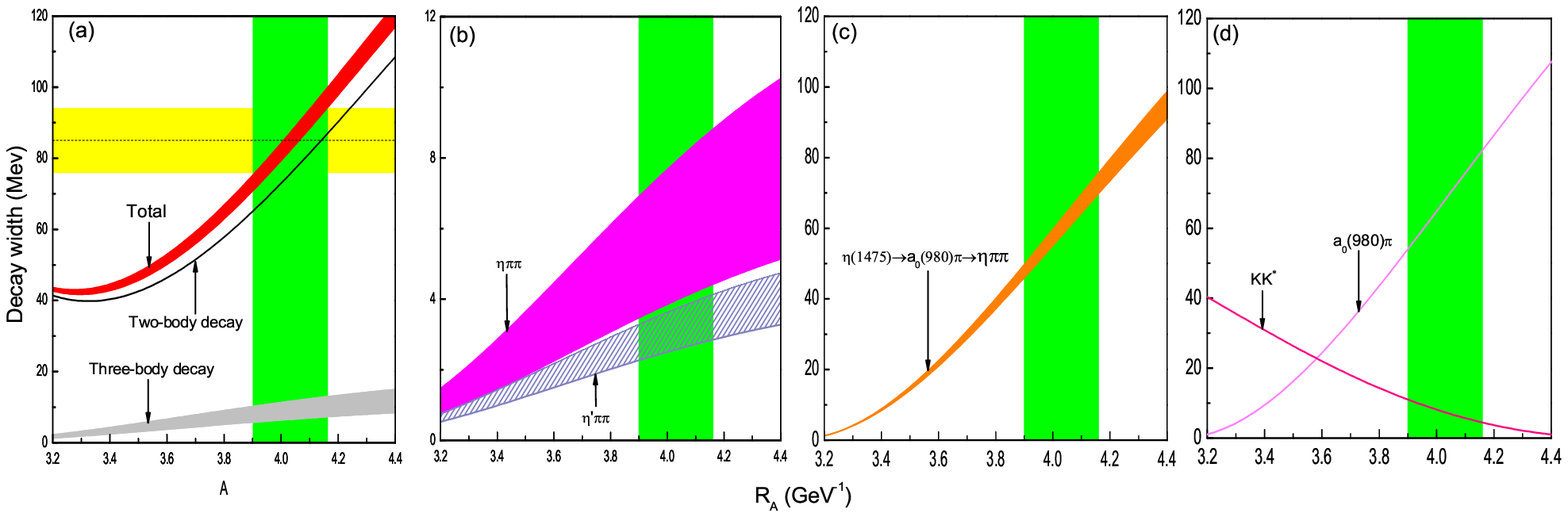}}\\
\end{tabular}
\caption{(Color online). The $R$ dependence of the calculated total
and partial widths of $\eta(1475)$. The partial widths presented in
(a)-(d) are arranged in the same way as in Fig.~\protect\ref{1295}.
}\label{1475}
\end{figure*}
\end{center}

The $R$-dependence of the total decay widths of
$\eta(1295)/\eta(1475)$ are shown in Figs. \ref{1295} (a) and
\ref{1475} (a), where their total decay widths are from the two-body
and double pion decays. For $\eta(1295)$, there exists overlap
between the theoretical result and experimental measurement with
$R=3.66\sim3.69$ GeV$^{-1}$, which is close to the value given in
Ref. \cite{Close:2005se}. For $\eta(1475)$, the $R$ range
($R=3.91\sim4.16$ GeV$^{-1}$) for the overlap between the
experimental and theoretical values is larger than that for
$\eta(1295)$ as shown in Fig. \ref{1475} (a).
Furthermore, with such a $R$ value for $\eta(1475)$, the partial
widths of $\eta(1475)\to a_0(980)\pi$ and $K\bar{K}^*+c.c.$ are
$55.1\sim82.3$ MeV and $10.7\sim4.4$ MeV, respectively. This seems
to contradict the experimental data since the dominant decay channel
of $\eta(1475)$ is $K\bar{K}\pi$, and $K\bar{K}\pi$ is possibly from
the decay of $K\bar{K}^*(892)+c.c.$ Such inconsistency can be due to
an unsuitable $R$ range for $\eta(1475)$. If taking the same $R$
range as that of $\eta(1295)$, we found the decay width of
$\eta(1475)\to K\bar{K}^*+c.c.$ is comparable with that of
$\eta(1475)\to a_0(980)\pi$, and the inconsistency mentioned above
does no longer exist. The obtained total decay width of $\eta(1475)$
($\Gamma=53.5\sim 58.7$ MeV) is smaller than the averaged value of
the width of $\eta(1475)$ ($\Gamma=85\pm 9$ MeV) listed in PDG
\cite{Amsler:2008zzb} and close to the central value ($\Gamma=54$
MeV) given by the Mark III Collaboration \cite{Bai:1990hs}.
Regarding this situation, we expect more precise experimental
measurement of the $\eta(1475)$ resonance parameter, e.g. from
BES-III, will help clarify this problem.

In Fig. \ref{1295} (b), it shows that $\eta(1295)$ can dominantly
decay into $\eta\pi\pi$ via the intermediate scalars. For
$\eta(1475)$, the $\eta\pi\pi$ decay width is a slightly larger than
that of $\eta^\prime \pi\pi$, as shown by Fig. \ref{1475} (b). Since
$\eta\pi$ is the dominant decay of $a_0(980)$ \cite{Amsler:2008zzb},
the $a_0(980)\pi$ should be an important contributing channel in
$\eta(1295)/\eta(1475)\to a_0(980)\pi\to \eta\pi\pi$. In Figs.
\ref{1295} (c) and  \ref{1475} (c), the $R$-dependences of the decay
widths of $\eta(1295)/\eta(1475)\to a_0(980)\pi\to \eta\pi\pi$ are
presented. We notice that the decay widths of
$\eta(1295)/\eta(1475)\to \eta\pi\pi$ via intermediate $a_0(980)$
are comparable with those via intermediate scalar states $\sigma$
and $f_0(980)$. It indicates that intermediate $a_{0}(980)$
contribution is important to the double pion decays of
$\eta(1295)/\eta(1475)$, especially to $\eta(1475)\to \eta\pi\pi$.
This can be tested by the experimental analysis of the $\eta\pi$ and
$\pi\pi$ invariant mass spectra of $\eta(1295)/\eta(1475)\to
\eta\pi\pi$.

\subsubsection{$\eta(1760)$ and $X(1835)$}

\begin{center}
\begin{figure*}[htbp]
\begin{tabular}{cc}
\scalebox{0.99}{\includegraphics{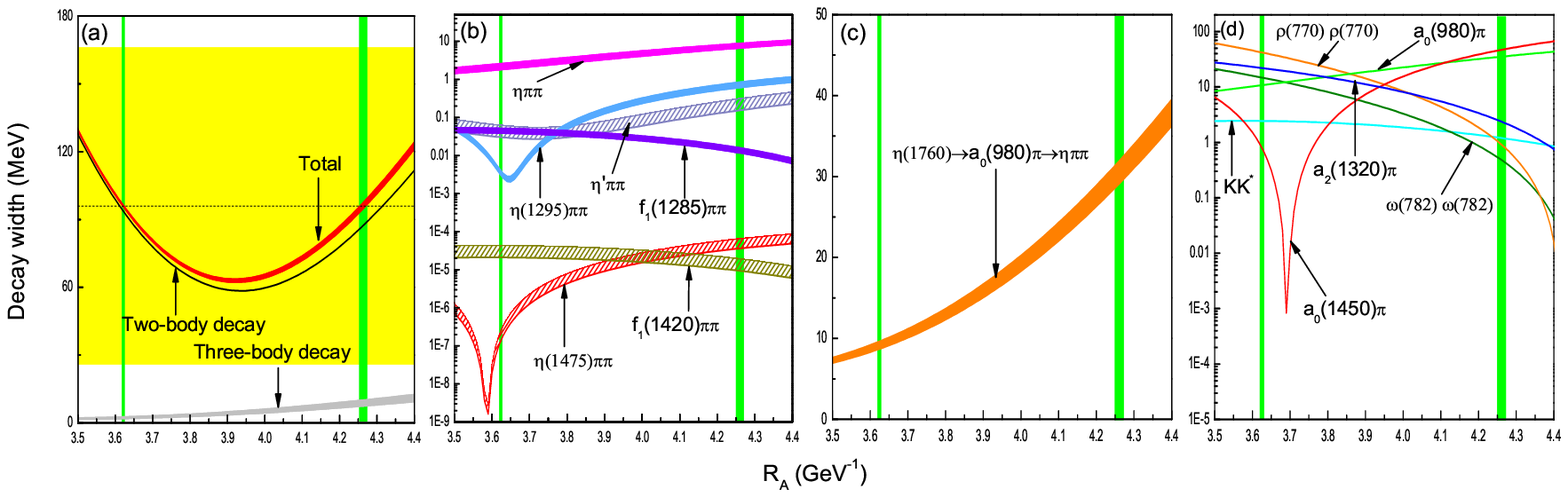}}
\end{tabular}
\caption{(Color online). The $R$ dependence of the calculated total
and partial widths of $\eta(1760)$. The partial widths presented in
(a)-(d) are arranged in the same way as in Fig.~\protect\ref{1295}.
}\label{1760}
\end{figure*}
\end{center}

\begin{center}
\begin{figure*}[htbp]
\begin{tabular}{cc}
\scalebox{0.95}{\includegraphics{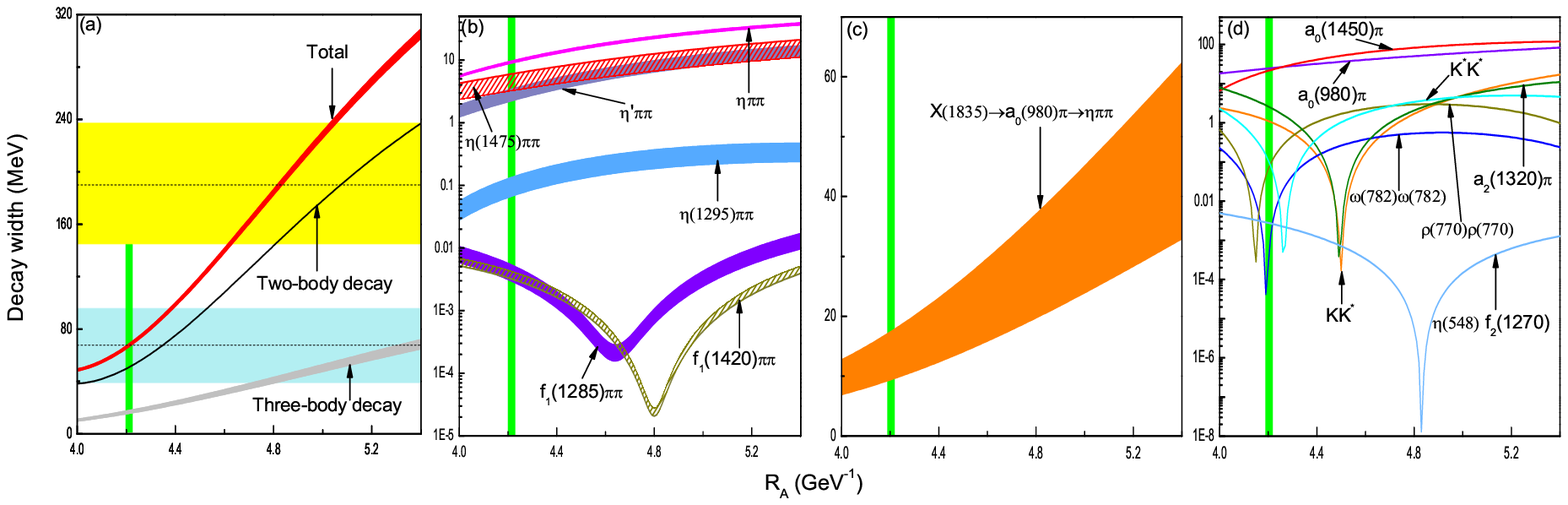}}
\end{tabular}
\caption{(Color online). The $R$ dependence of the calculated total
and partial widths of $X(1835)$. The partial widths presented in
(a)-(d) are arranged in the same way as in Fig.~\protect\ref{1295}.
Here, dashed lines with yellow and blue bands are the BES-II \cite{Ablikim:2005um} and BES-III \cite{Collaboration:2010au} measurements of $X(1835)$ width, respectively.
}\label{1835}
\end{figure*}
\end{center}

The calculation results for the $\eta(1760)$ and $X(1835)$ decays in
terms of $R$ are presented in Figs.  \ref{1760} and \ref{1835},
respectively. As shown by Figs.  \ref{1760} (a) and \ref{1835} (a),
the total decay width of $\eta(1760)$ can fit the central value of
the experimental width with $R=3.62\sim3.63$ GeV$^{-1}$ or
$R=4.26\sim4.27$ GeV$^{-1}$, due to large experimental uncertainties
with the $\eta(1760)$ width.

To some extent, the behavior of partial decay width of $\eta(1760)$
with $R=3.62\sim3.63$ GeV$^{-1}$ (see Fig. \ref{1760} (d)) can
reflect the experimental observation of $\eta(1760)$ well. In this
range, the partial widths are consistent with its signals observed
in $J/\psi\to \gamma \omega\omega$ \cite{Ablikim:2006ca} and
$\gamma\rho\rho$ \cite{Bisello:1988as}. The two-body partial decay
widths of $\eta(1760)$ are given in Fig. \ref{1760} (d), from which
one sees that $\rho(770)\rho(770)$, $a_2(1320)\pi$,
$\omega(782)\omega(782)$ and $a_0(980)\pi$ are the main decay
channels for $\eta(1760)$, i.e. $\Gamma(\eta(1760)\to
\rho(770)\rho(770))=41.3\sim42.7$ MeV, $\Gamma(\eta(1760)\to
a_2(1320)\pi)=21.8\sim22.3$ MeV, $\Gamma(\eta(1760)\to
\omega(782)\omega(782))=15.0\sim14.6$ MeV and $\Gamma(\eta(1760)\to
a_0(980)\pi)=10.6\sim10.8$ MeV.

BES-II and BES-III have given different  widths (the dashed lines
with yellow and blue bands) for $X(1835)$ as shown in Fig.
\ref{1835} (a). When comparing our calculation with the experimental
width, we still adopt the resonance parameter from BES-II. It shows
that the calculated decay width of $X(1835)$ under the assignment of
the second radial excitation of $\eta^\prime(958)$ is consistent
with the BES-II measurement. The obtained $R$ falls into
$4.20\sim4.22$ GeV$^{-1}$, which corresponds to the width of
$X(1835)$ given by BES-II. This $R$ value is quite close to that for
$\eta(1760)$, and consistent with the estimate in Ref.
\cite{Close:2005se}. The results presented in Fig. \ref{1835} (d)
further indicate that the decay widths of $X(1835)$ into
$a_0(980)\pi$ and $a_0(1450)\pi$ can reach up to $24.5\sim25.2$ MeV
and $21.3\sim22.2$ MeV, respectively.

The partial decay widths obtained by the QPC model seem to support
the explanation for $\eta(1760)$ and $X(1835)$ as  the second radial
excitations of $\eta(548)$ and $\eta^\prime(958)$. For the double
pion decays, we find that the intermediate scalar productions appear
to be the main contributor as shown in Figs. \ref{1760} (b) and
\ref{1835} (b).



Due to $\eta\pi$ being the dominant decay channel of $a_0(980)$,  in
Figs. \ref{1760} (c) and \ref{1835} (c) the $R$ dependence of the
decay widths of $\eta(1760)/X(1835)\to a_0(980)\pi\to \eta\pi\pi$
also shows that contributions from intermediate $a_0$ productions
are comparable with those via intermediate $\sigma$ and $f_0(980)$
in $\eta(1760)/X(1835)\to \eta\pi\pi$. This prediction can be
checked in future experiment.

\subsubsection{$X(2120)$}

\begin{center}
\begin{figure*}[htbp]
\begin{tabular}{cc}
\scalebox{1}{\includegraphics{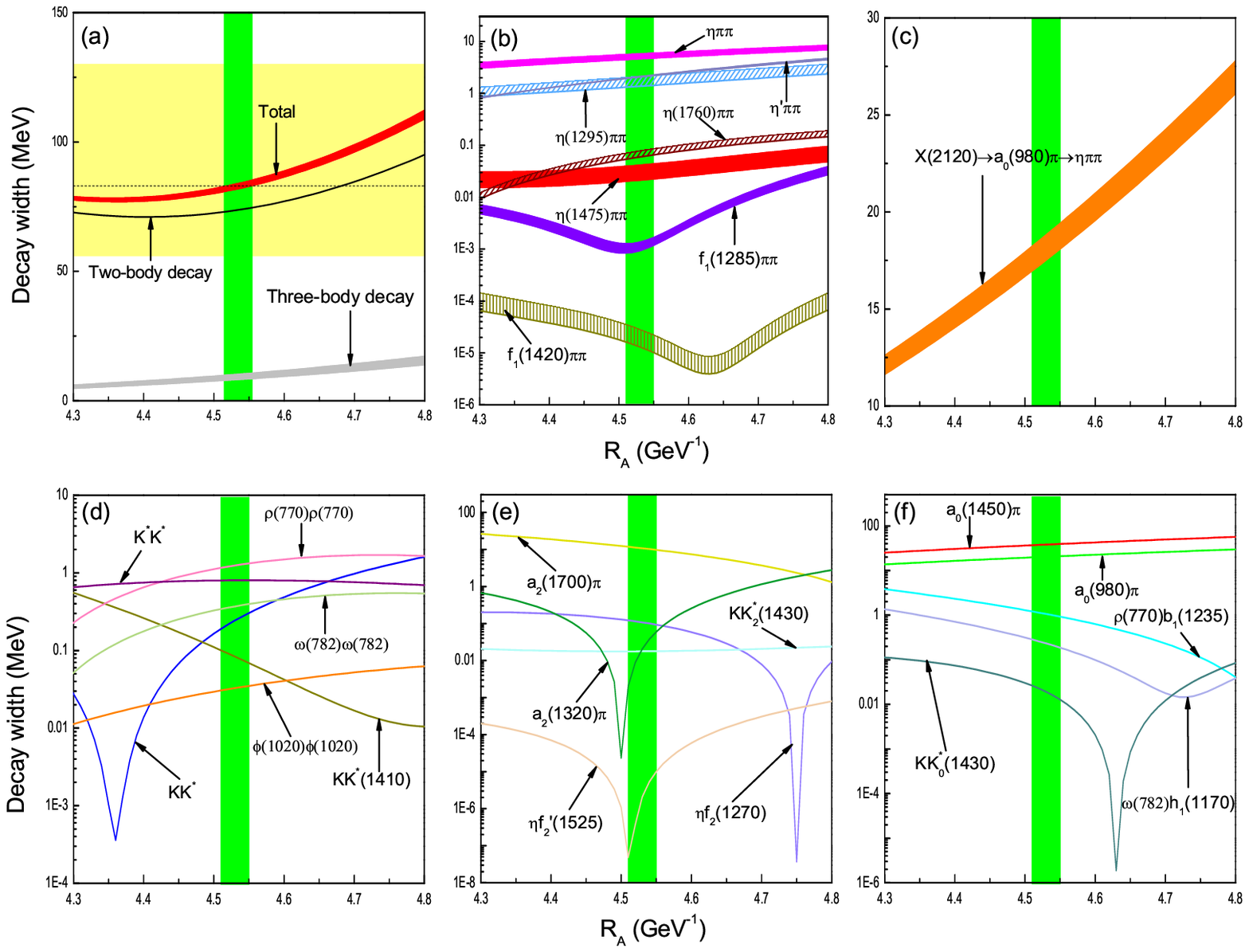}}
\end{tabular}
\caption{(Color online). The $R$ dependence of the calculated total
and partial widths of  $X(2120)$  as the third radial excitation of
$\eta(548)$. The partial widths presented in (a)-(c) are arranged in
the same way as in Fig.~\protect\ref{1295}, while the  partial
two-body decay widths are given in (d)-(f).
%
}\label{2120eta}
\end{figure*}
\end{center}

\begin{center}
\begin{figure*}[htbp]
\begin{tabular}{cc}
\scalebox{1}{\includegraphics{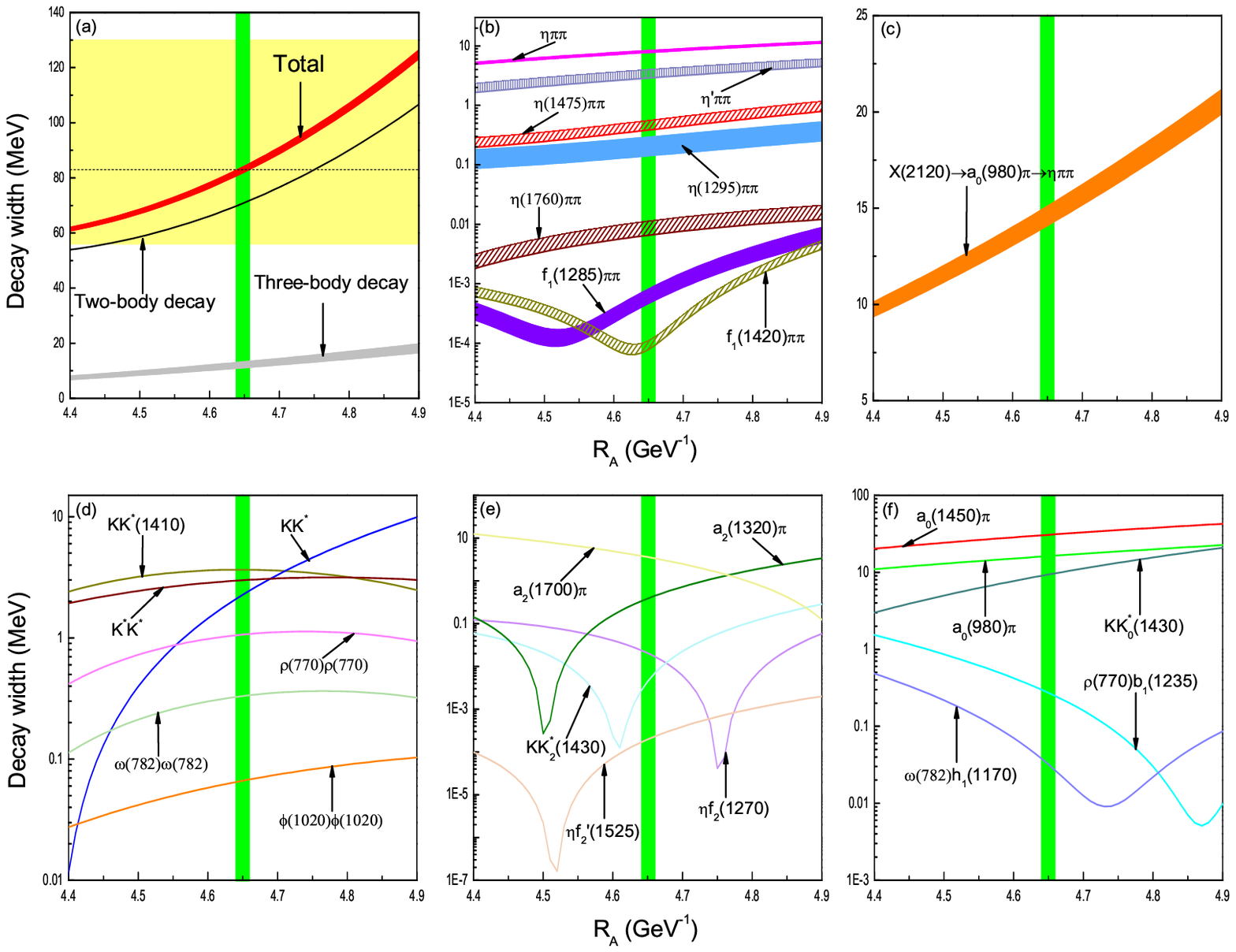}}
\end{tabular}
\caption{(Color online). The $R$ dependence of the calculated total
and partial widths of  $X(2120)$  as the third radial excitation of
$\eta(958)$. The partial widths presented in (a)-(c) are arranged in
the same way as in Fig.~\protect\ref{1295}, while the  partial
two-body decay widths are given in (d)-(f).
}\label{2120etaprime}
\end{figure*}
\end{center}

The strong decay behaviors of $X(2120)$ are presented in Figs.
\ref{2120eta}-\ref{2120etaprime}. As shown in Fig. \ref{2120eta}
(a),  the experimental width of $X(2120)$ can be reproduced by
assuming $X(2120)$ as the third radial excitation of $\eta(548)$
with $R=4.51\sim4.55$ GeV$^{-1}$. This range is also consistent with
the $R$ range estimated in Ref. \cite{Close:2005se}. The two-body
partial decay widths of $X(2120)$ in Fig. \ref{2120eta} (d)-(f)
indicate that $a_0(1450)\pi$, $a_0(980)\pi$ and $a_2(1700)\pi$ are
the dominant decay channels, i.e. $\Gamma(X(2120)\to
a_0(1450)\pi)=37.6\sim40.1$ MeV, $\Gamma(X(2120)\to
a_0(980)\pi)=20.0\sim21.3$ MeV and $\Gamma(X(2120)\to
a_2(1700)\pi)=9.2\sim11.3$ MeV. It should be noted that decays
$X(2120)\to a_0(1450)\pi, a_0(980)\pi$ occur via $S$-wave while
$X(2120)\to a_2(1700)\pi$ is via $P$-wave, which explains that the
$P$-wave decay width is smaller than the $S$-wave.

The decay of $X(2120)\to a_0(980)\pi\to \eta\pi\pi$ occurs  with
considerable decay width due to the large coupling of $a_0(980)\to
\eta\pi$. The results are shown in Fig. \ref{2120eta} (c). Moreover,
 intermediate scalar states, i.e.
$\sigma$ and $f_0(980)$, have also important contributions to the
double pion decay channels, e.g.  $X(2120)\to \eta\pi\pi$,
$\eta^\prime\pi\pi$. The results are presented in Fig. \ref{2120eta}
(b).

Under the assignment of the third radial excitation  of
$\eta^\prime(958)$, one obtains the total decay width of $X(2120)$
(see Fig. \ref{2120etaprime} (a)). We can still find the overlap
between theoretical and experimental total decay widths, with the
obtained central values of $R= 4.64\sim 4.66$ GeV$^{-1}$. This range
is consistent with that for $X(2120)$ as the third radial excitation
of $\eta(548)$. The results in Fig. \ref{2120etaprime} (d)-(f) show
that $a_0(1450)\pi$ and $a_0(980)\pi$ are the main decay modes of
$X(2120)$ if it is the third radial excitation of $\eta(958)$. This
determines a sizeable decay width of $X(2120)\to a_0(980)\pi\to
\eta\pi\pi$ (see Fig. \ref{2120etaprime} (c)). Again, we find
important contributions from the intermediate scalar state in the
double pion decays (see Fig. \ref{2120etaprime} (b)). The
calculation suggests that both $\eta\pi\pi$ and $\eta^\prime\pi\pi$
are important double pion decay modes of $X(2120)$.


Note that in Sec. \ref{sec2}, the analysis of the mass spectrum of
$\eta-\eta^\prime$ family indicates that there exists the partner of
$X(2120)$ with the mass very close to that of $X(2120)$. Because of
this, it is natural to set the mass of the partner of $X(2120)$ the
same as 2.12 GeV. Thus, the above results should have reflected the
theoretical expectations of the $X(2120)$ decay behavior as the
third radial excitation of $\eta(548)$ or $\eta^\prime(958)$.

\subsubsection{$X(2370)$}

\begin{center}
\begin{figure*}[htbp]
\begin{tabular}{cc}
\scalebox{0.9}{\includegraphics{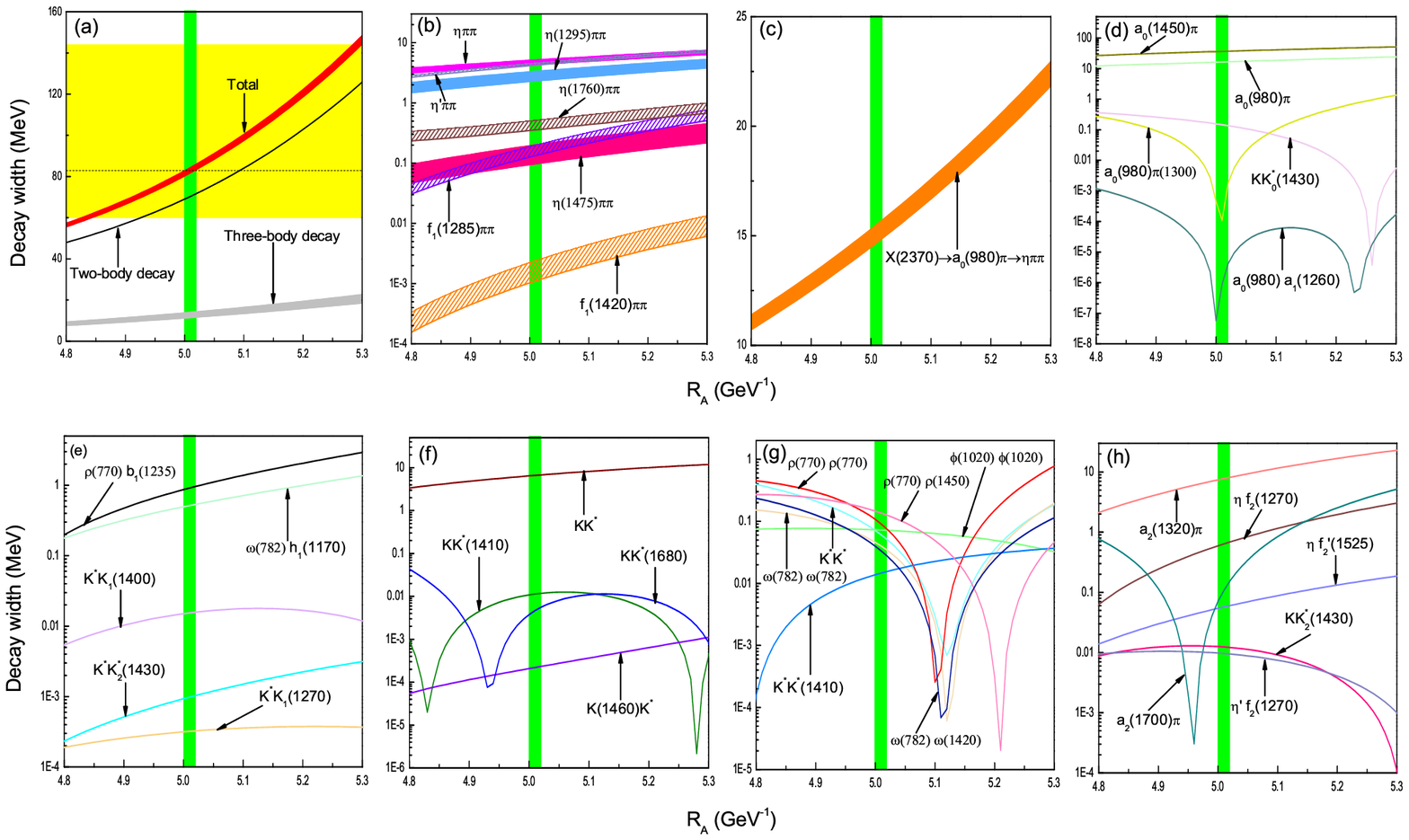}}
\end{tabular}
\caption{(Color online). The $R$ dependence of the calculated total
and partial widths of  $X(2370)$  as the fourth radial excitation of
$\eta(548)$. The partial widths presented in (a)-(c) are arranged in
the same way as in Fig.~\protect\ref{1295}, while the  partial
two-body decay widths are given in (d)-(f).
}\label{2370eta}
\end{figure*}
\end{center}

\begin{center}
\begin{figure*}[htbp]
\begin{tabular}{c}
\scalebox{0.9}{\includegraphics{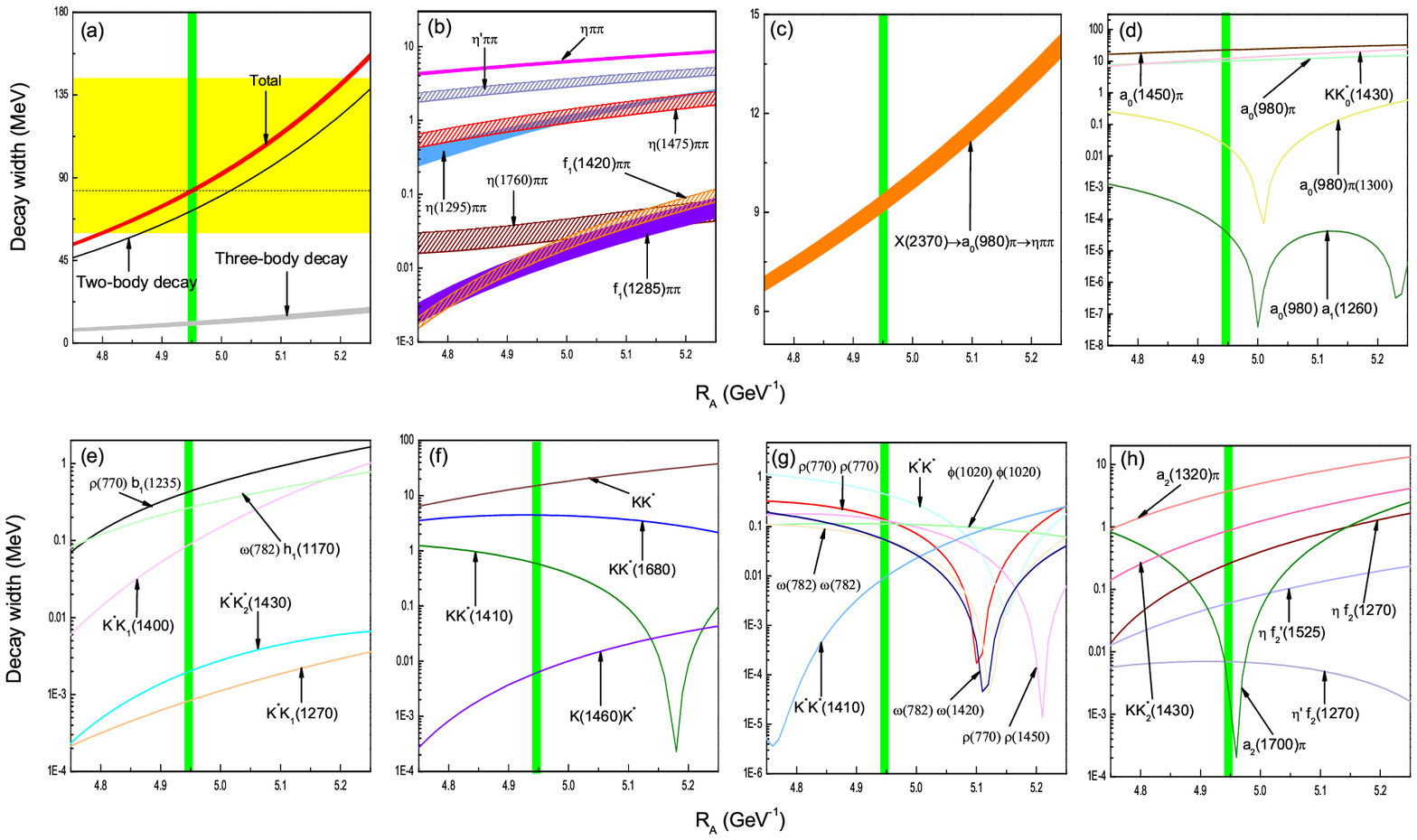}}
\end{tabular}
\caption{(Color online). The $R$ dependence of the calculated total
and partial widths of  $X(2370)$  as the fourth radial excitation of
$\eta(958)$. The partial widths presented in (a)-(c) are arranged in
the same way as in Fig.~\protect\ref{1295}, while the  partial
two-body decay widths are given in (d)-(f).
}\label{2370etaprime}
\end{figure*}
\end{center}

In this Subsection, we present the results for $X(2370)$ as the
fourth radial excitation of $\eta(548)$ or $\eta^\prime(958)$ in
Figs. \ref{2370eta} and \ref{2370etaprime}. Notice that more decay
channels are open for $X(2370)$.

As the fourth radial excitation of $\eta(548)$,  the obtained total
decay width of $X(2370)$ are consistent with the experimental data
(see Fig. \ref{2370eta} (a)), where the central values of $R$ are
about 5 GeV$^{-1}$. The corresponding two-body decays indicates
$X(2370)$ mainly decays into $a_0(980)\pi$, $a_0(980)\pi$, which are
shown in Fig. \ref{2370eta} (d)-(h). In addition, $KK^*$ and
$a_2(1320)\pi$ channels also play an important role here. These
implies the dominant contributions from intermediate scalars to the
double pion decays of $X(2370)\to \eta\pi\pi$, $\eta^\prime\pi\pi$
and $\eta(1295)\pi\pi$ as shown in  Fig. \ref{2370eta} (c).


If categorizing $X(2370)$ as the fourth radial excitation of
$\eta^\prime(958)$, one also obtains the total decay width of
$X(2370)$ consistent with the experimental data with
$R=4.95\sim4.96$ GeV$^{-1}$. The comparison is shown in Fig.
\ref{2370etaprime} (a). The results in Fig. \ref{2370etaprime}
(d)-(h) provide valuable information of main decay channels of
$X(2370)$. Several main decay channels of $X(2370)$ include
$a_0(1450)\pi$, $KK^*$, $KK^*_0(1430)$, $a_0(980)\pi$, $KK^*(1680)$
and $a_2(1320)\pi$. We list their partial decay widths as a
comparison: $\Gamma(X(2370)\to a_0(1450)\pi)=22.6\sim22.9$ MeV,
$\Gamma(X(2370)\to KK^*)=15.4\sim15.9$ MeV, $\Gamma(X(2370)\to
KK_0^*(1430))=12.2\sim12.5$ MeV, $\Gamma(X(2370)\to
a_0(980)\pi)=10.1\sim10.3$ MeV, $\Gamma(X(2370)\to KK^*(1680))=4.4$
MeV and $\Gamma(X(2370)\to a_2(1320)\pi)=3.8\sim4.0$ MeV. Again, it
shows that the double pion decay widths will have important
contributions from the intermediate scalars as shown in Fig.
\ref{2370etaprime} (b)


Note that $X(2370)\to KK_3^*(1780)$ is a $F$-wave decay, thus, will
be suppressed in comparison with other low partial wave decays. In
Figs. \ref{2370eta} and \ref{2370etaprime} we do not include the
result of $X(2370)\to KK_3^*(1780)$.

\hspace{1cm}

\begin{center}
\begin{figure*}[htbp]
\begin{tabular}{cccccc}
\scalebox{0.6}{\includegraphics{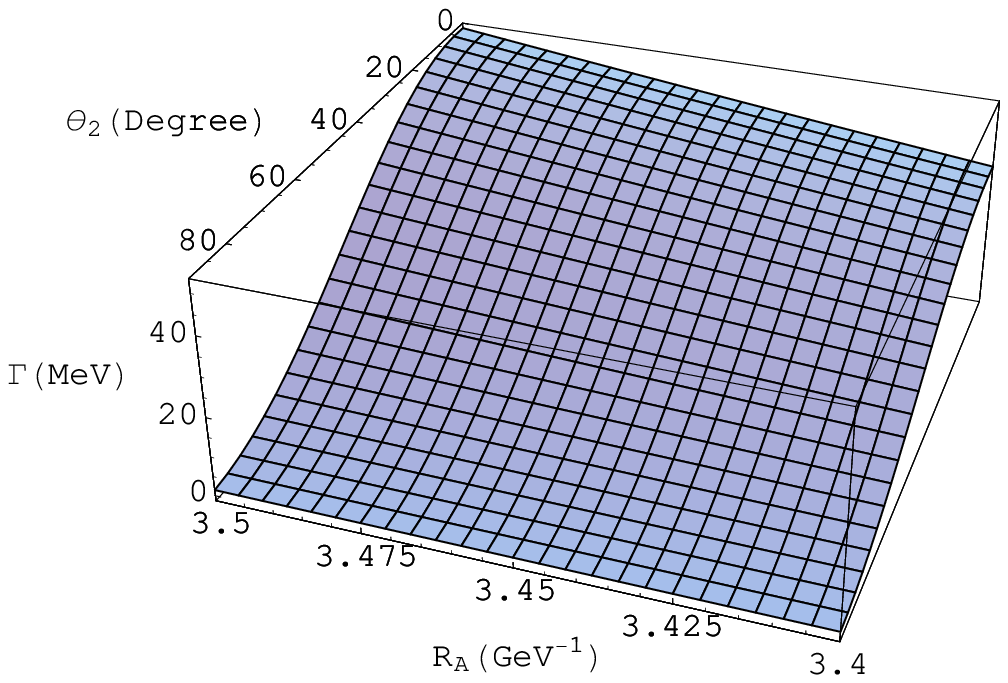}}&\scalebox{0.6}{\includegraphics{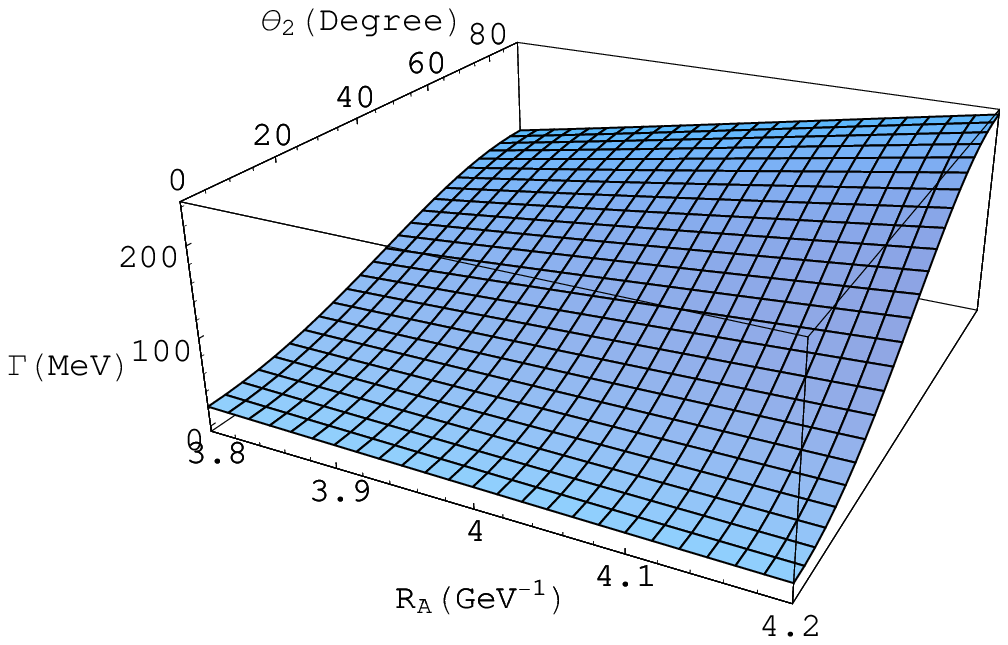}}\\
(a)&(b)\\
\scalebox{0.6}{\includegraphics{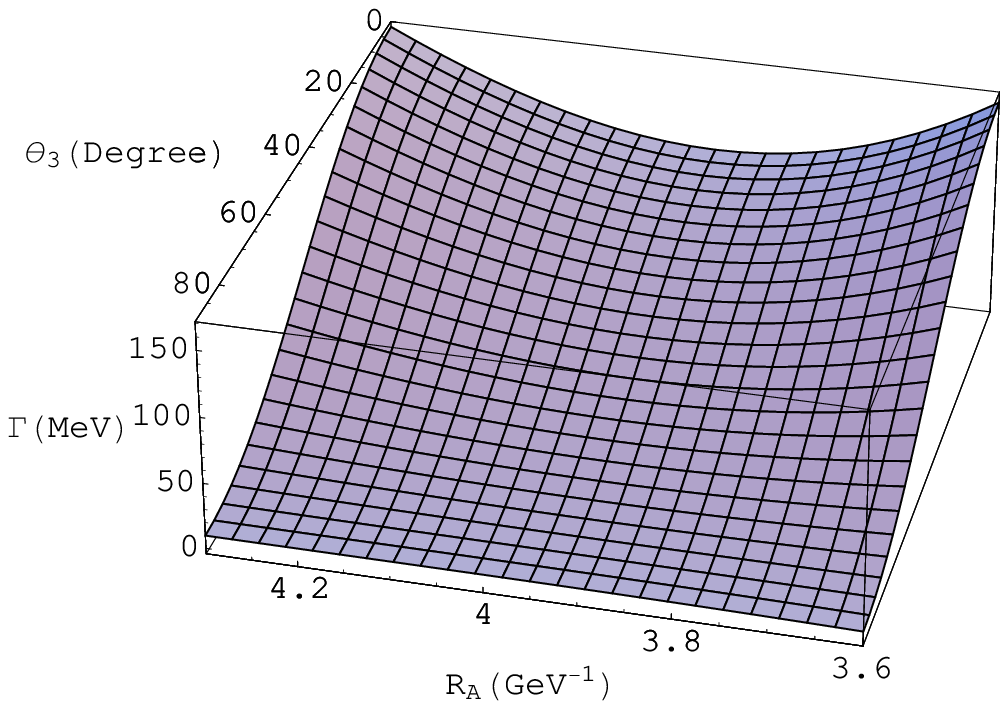}}&\scalebox{0.6}{\includegraphics{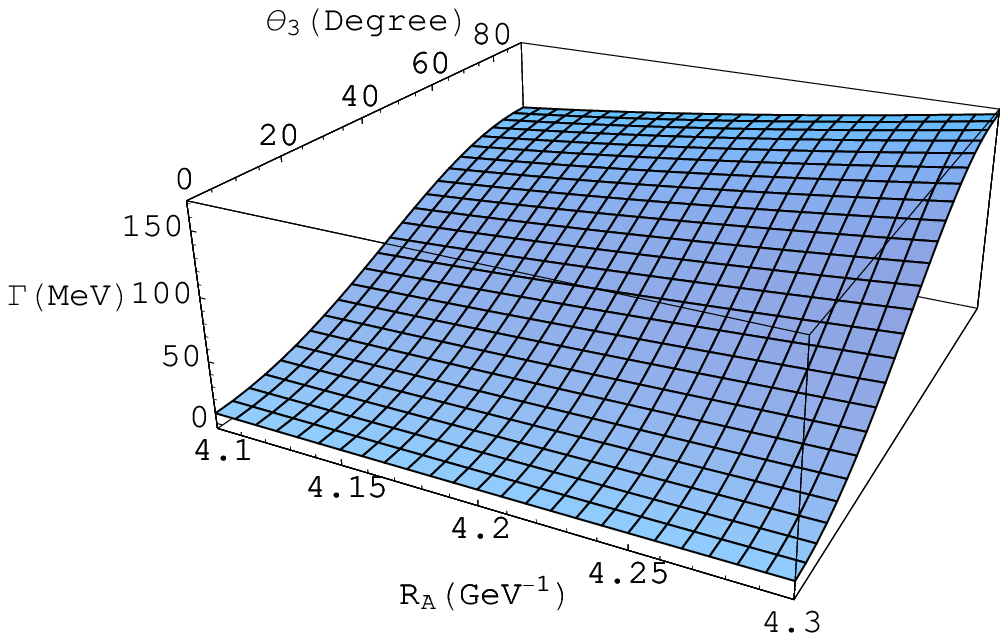}}\\
(c)&(d)\\
\scalebox{0.6}{\includegraphics{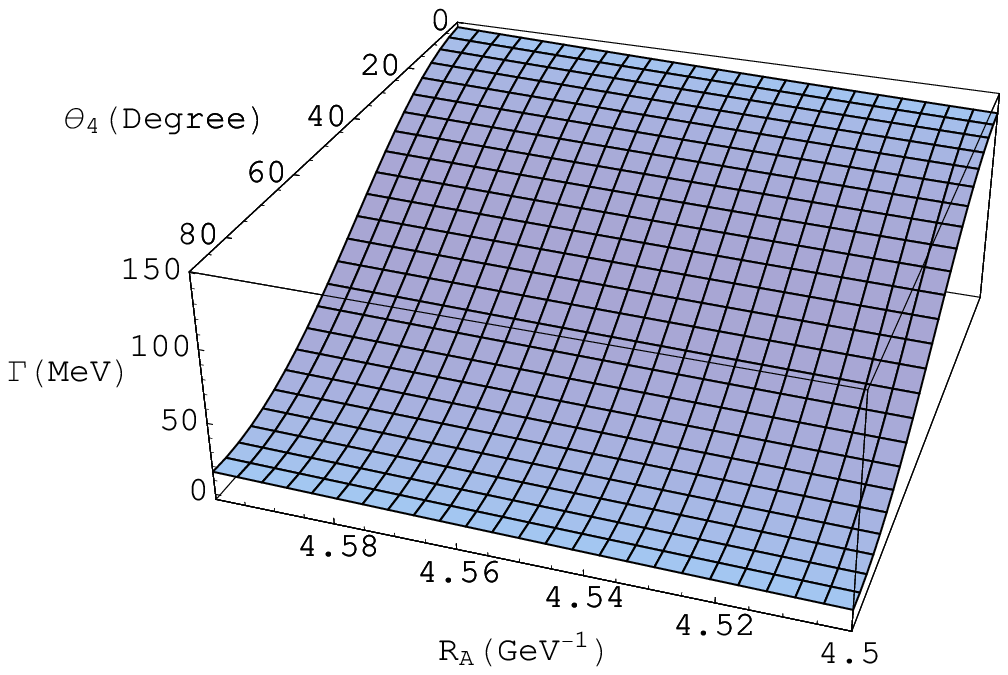}}&\scalebox{0.6}{\includegraphics{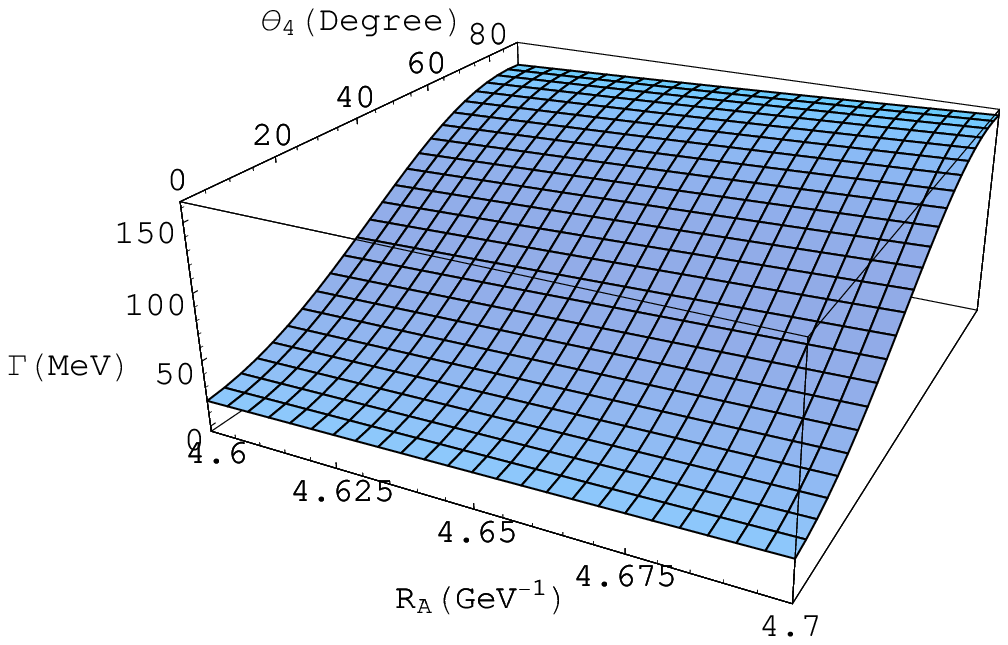}}\\
(e)&(f)\\
\scalebox{0.6}{\includegraphics{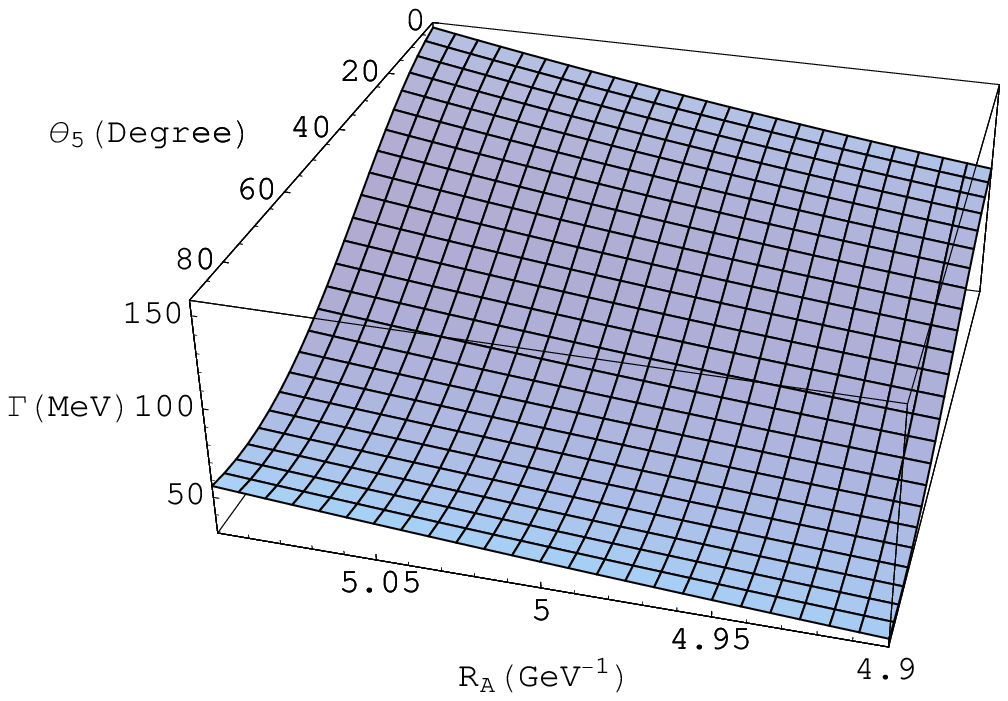}}&\scalebox{0.6}{\includegraphics{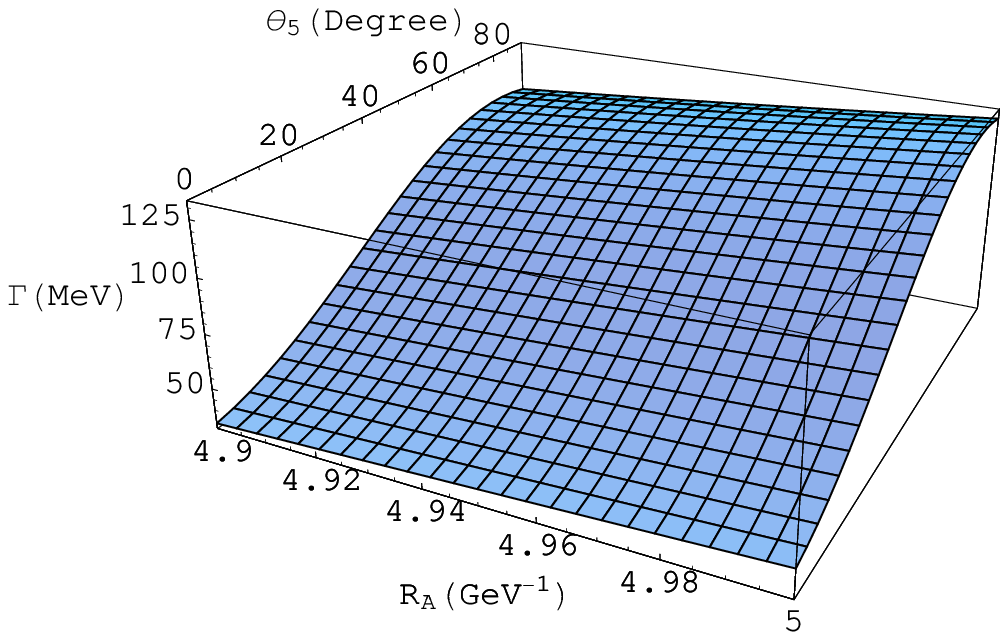}}\\
(g)&(h)
\end{tabular}
\caption{(Color online).  The $R$ and $\theta_i$ ($i=2,\cdots,5$)
dependence of the total decay widths of $\eta(1295)$, $\eta(1475)$,
$\eta(1760)$, $X(1835)$, $X(2120)$ and $X(2370)$. (a), (b), (c) and
(d) correspond to the decays of $\eta(1295)$, $\eta(1475)$,
$\eta(1760)$ and $X(1835)$ respectively. Diagrams (e)/(f) shows the
decay behaviors of $X(2120)$ as the third radial excitation of
$\eta(548)/\eta^\prime(958)$, while the variation of the decay
behaviors of $X(2370)$ to $\theta_5$ and $R$ is given in diagrams
(g)/(h) with $X(2370)$ as the fourth radial excitation of
$\eta(548)/\eta^\prime(958)$. }\label{3D}
\end{figure*}
\end{center}

We need to specify that the above numerical results are obtained by
taking the mixing angle of $\eta(1295)/\eta(1475)$,
$\eta(1760)/X(1835)$, $X(2120)$ and $X(2370)$ the same as the
$\eta(548)/\eta'(958)$ mixing angle, i.e.
$\theta_5=\theta_4=\theta_3=\theta_2=\theta_1=39.3^\circ$. For
further studying the total decay widths of $\eta(1295)/\eta(1475)$,
$\eta(1760)/X(1835)$, $X(2120)$ and $X(2370)$ dependent on
$\theta_i$ and $R$, in Fig. \ref{3D} we present the 3D plot of the
total decay widths of $\eta(1295)/\eta(1475)$, $\eta(1760)/X(1835)$,
$X(2120)$ and $X(2370)$ with variables $\theta_i$ ($i=1,\cdots,5$)
and $R$. It would be useful for further constraining the theoretical
calculations by experimental measurements.

\section{Discussion and conclusion}\label{sec5}

The confirmation of $X(1835)$ and  observation of two new
resonances, $X(2120)$ and $X(2370)$, by the BES-III Collaboration
have inspired a lot efforts in the study of light hadron
spectroscopy. In particular, it largely enriches our knowledge about
the isoscalar and pseudoscalar spectrum. As stated earlier,
$\pi(1300)$, $K(1460)$, $\eta(1295)$, and $\eta(1475)$ naturally
make the first radial excitation states of $0^-$ mesons. Our
speculation is that the second radial excitation nonet can be formed
by $\pi(1800)$, $K(1830)$, $\eta(1760)$ and $X(1835)$. This
categorizing scheme is in good agreement with the qualitative
expectation of their mass spectrum in the constituent quark model.
This scheme also allows us to accommodate $X(2120)$ and $X(2375)$ in
the isoscalar pseudoscalar family as higher radial excitation
states.

Within this categorizing scheme, we calculate the strong decay
widths of these states in the QPC model to compare with the
available experimental data. The results show that $\eta(1295)$,
$\eta(1760)$ are good candidates of the first and the second radial
excitations of $\eta(548)$, while $X(1835)$ can be explained as the
second radial excitation of $\eta'(958)$. With a reasonable $R$
range, the calculated total decay widths  of $\eta(1295)$,
$\eta(1760)$, $X(1835)$ agree well with the experimental data.

For $\eta(1475)$, the $R$ range is larger than that for $\eta(1295)$
if one requires the calculated width to fit the experimental data
well. Under such a condition, the decay mode $a_0(98)\pi$ becomes
the main contributor to the total width for $\eta(1475)$ which,
however, does not coincide with the experimental observation since
$\eta(1475)$ mainly decays into $K\bar{K}\pi$. By assigning
$\eta(1475)$ as the partner of $\eta(1295)$, and thus its $R$ range
being the same as that of $\eta(1295)$, we find that the decay width
of $K\bar{K}^*+c.c.$ becomes comparable with that of $a_0(980)\pi$.
The dominance of the $K\bar{K}\pi$ decay mode can be recovered. It
is likely that the total width of $\eta(1475)$ was overestimated by
experiment due to large uncertainties. Therefore, more precise data
for the $\eta(1475)$ resonance parameters are strongly recommended
in future experiment.

Our calculation suggests  that $X(2120)$ and $X(2370)$ could be as
the third and fourth radial excitations of
$\eta(548)/\eta^\prime(958)$, respectively. Since the masses of the
partners of the same radial excitations are close to each other, we
analyze both possibilities and predict their strong decay patterns.
It should be useful for experimental search for their partners in
other channels such as $\eta\pi\pi$ and $K\bar{K}\pi$.

It is interesting to note that
apart from those three enhancements $X(1835)$, $X(2120)$ and
$X(2370)$, an enhancement around 1.5 GeV also appears in the
$\eta^\prime\pi\pi$ invariant spectrum~\cite{Collaboration:2010au}.
A possible assignment to $f_1(1510)$ was discussed in Ref.
\cite{Collaboration:2010au}. We notice that that a pseudoscalar
state $\eta(1475)$ would be favored by an $S$-wave decay into
$\eta^\prime \pi\pi$, while $f_1(1510)\to \eta^\prime \pi\pi$ would
be via $P$-wave. Our calculation shows that $\eta(1475)$ should have
a sizeable contribution to the $\eta'\pi\pi$ invariant mass
spectrum. The slight mass shift could be due to its interferences
with other contributions. Thus, to make sure wether there is $\eta(1475)$ contribution in the $\eta^\prime \pi^+\pi^-$ mode, a partial wave
analysis with $\eta(1475)$ included in the fit should be tested.
We also mention that, as the second radial
excitation of $\eta(548)$, the dominant double pion decay channel of
$\eta(1760)$ is via $\eta(1760)\to \eta \pi\pi$ instead of
$\eta(1760)\to \eta^\prime \pi\pi$. Therefore, it may not appear
predominant in the $\eta^\prime\pi\pi$ invariant mass spectrum. It
should also be recognized that the inclusion of $\eta(1760)$ will
have important interfering effects with other resonance amplitudes,
which could be essential for extracting the resonance parameters in
the partial wave analysis. We expect that the future partial wave
analysis from BES-III can testify this point.

To summarize, the new data from BES-III have provided important
information on the pseudoscalar spectrum, and could be so-far the
first observation of higher radial excitations of $\eta/\eta'$
states. It will greatly enrich our knowledge about the light hadron
spectrum, and be useful for further efforts on the study of exotic
states in both experiment and theory. We expect that more
experimental data from BES-III in the near future will bring great
opportunities for our understanding of the strong QCD in the light
hadron sector.

\section*{Acknowledgements}

We would like thank Shan Jin, Xue-Qian Li, Hai-Yang Cheng, Dian-Yong Chen, Jun He, Bin Chen and D.~V. Bugg
for useful discussions. This project is supported in part by the
National Natural Science Foundation of China (Grants No. 11035006,
No. 11047606, No. 11035006), the Ministry of Education of China
(FANEDD under Grants No. 200924, DPFIHE under Grants No.
20090211120029, NCET under Grants No. NCET-10-0442, the Fundamental
Research Funds for the Central Universities under Grants No.
lzujbky-2010-69), and the Chinese Academy of Sciences
(KJCX2-EW-N01).

\begin{table*}[htbp]
\begin{tabular}{c|c|c|c|c|c}\toprule[1pt]
Particle&Mass (MeV)&Flavor wave function&$J^{PC}$&$n\;^{2s+1}L_J$&$R$ (GeV$^{-1}$) \cite{Close:2005se}\\\midrule[1pt]
$\eta$ \cite{Feldmann:1998vh}&548&$\cos\theta_1(\frac{u\bar{u}+d\bar{d}}{\sqrt{2}})-\sin\theta_1({s\bar{s}}),\,\theta_1=39.3^\circ$&$0^{-+}$&$1\;^{1}S_0$&2.106\\
$\sigma$ \cite{Cheng:2002ai,Anisovich:2001zp}
&600&$-\sin\varphi(s\bar{s})+\cos\varphi(\frac{u\bar{u}+d\bar{d}}{\sqrt{2}}),\,\varphi=140^\circ$&$0^{++}$&$1\;^{3}P_0$&3.486\\
$\rho$ \cite{Amsler:2008zzb}&775.49&$\rho^+=u\bar{d},\,\rho^-=\bar{u}d,\,\rho^0=\frac{u\bar{u}-d\bar{d}}{\sqrt{2}}$&$1^{--}$&$1\;^{3}S_1$&3.571\\
$\omega $&782.65&$\frac{u\bar{u}+d\bar{d}}{\sqrt{2}}$&$1^{--}$&$1\;^{3}S_1$&3.571\\
$\eta\prime$ \cite{Feldmann:1998vh}&958&$\sin\theta_1(\frac{u\bar{u}+d\bar{d}}{\sqrt{2}})+\cos\theta_1({s\bar{s}}),\,\theta_1=39.3^\circ$&$0^{-+}$&$1\;^{1}S_0$&2.106\\
$f_0(980)$ \cite{Cheng:2002ai,Anisovich:2001zp}
&980&$\cos\varphi(s\bar{s})+\sin\varphi(\frac{u\bar{u}+d\bar{d}}{\sqrt{2}}),\,\varphi=140^\circ$&$0^{++}$&$1\;^{3}P_0$&3.486\\
$a_0(980)$ \cite{Amsler:2008zzb}&984.7&$a_0^+=u\bar{d},\,a_0^-=\bar{u}d,\,a_0^0=\frac{u\bar{u}-d\bar{d}}{\sqrt{2}}$&$0^{++}$&$1\;^{3}P_0$&3.846\\
$\phi(1020) $&1019.45&${s\bar{s}}$&$1^{--}$&$1\;^{3}S_1$&2.778\\

$h_1(1170)$ \cite{Li:2005eq}&1170&$0.997(\frac{u\bar{u}+d\bar{d}}{\sqrt{2}})+0.073({s\bar{s}})$&$1^{+-}$&$1\;^{1}P_1$&3.704\\
$b_1(1235)$ \cite{Amsler:2008zzb}&1229.5&$b_1^+=u\bar{d},\,b_1^-=\bar{u}d,\,b_1^0=\frac{u\bar{u}-d\bar{d}}{\sqrt{2}}$&$1^{+-}$&$1\;^{1}P_1$&3.704\\
$a_1(1260)$ \cite{Amsler:2008zzb}&1230&$a_1^+=u\bar{d},\,a_1^-=\bar{u}d,\,a_1^0=\frac{u\bar{u}-d\bar{d}}{\sqrt{2}}$&$1^{++}$&$1\;^{3}P_1$&3.846\\
$f_2(1270)$ \cite{Li:2000zb,Geng:2009iw}&1275&$\frac{u\bar{u}+d\bar{d}}{\sqrt{2}}$&$2^{++}$&$1\;^{3}P_2$&3.846\\
$f_1(1285)$ \cite{Kirchbach:1995ep}&1281.8&$\cos\epsilon(\frac{u\bar{u}+d\bar{d}}{\sqrt{2}})-\sin\epsilon({s\bar{s}}),\,\epsilon=35.3^\circ$&$1^{++}$&$1\;^{3}P_1$&3.486\\
$\eta(1295)$&1295&$\cos\theta_2(\frac{u\bar{u}+d\bar{d}}{\sqrt{2}})-\sin\theta_2({s\bar{s}}),\,\theta_2=?$&$0^{-+}$&$2\;^{1}S_0$&3.301\\
$\pi(1300)$ \cite{Amsler:2008zzb}&1300&$\pi^+=u\bar{d},\,\pi^-=\bar{u}d,\,\pi^0=\frac{u\bar{u}-d\bar{d}}{\sqrt{2}}$&$0^{-+}$&$2\;^{1}S_0$&3.571\\
$a_2(1320)$ \cite{Amsler:2008zzb}&1318.3&$a_2^+=u\bar{d},\,a_2^-=\bar{u}d,\,a_2^0=\frac{u\bar{u}-d\bar{d}}{\sqrt{2}}$&$2^{++}$&$1\;^{3}P_2$&3.846\\

$f_1(1420)$ \cite{Kirchbach:1995ep}&1426.4&$\sin\epsilon(\frac{u\bar{u}+d\bar{d}}{\sqrt{2}})+\cos\epsilon({s\bar{s}}),\,\epsilon=35.3^\circ$&$1^{++}$&$1\;^{3}P_1$&3.486\\
$\omega(1420)$ \cite{Amsler:2008zzb}&$1400\sim1450$&$\frac{u\bar{u}+d\bar{d}}{\sqrt{2}}$&$1^{--}$&$2\;^{3}S_1$&4.167\\
$\rho(1450)$ \cite{Amsler:2008zzb}&1465&$\rho^+=u\bar{d},\,\rho^-=\bar{u}d,\,\rho^0=\frac{u\bar{u}-d\bar{d}}{\sqrt{2}}$&$1^{--}$&$2\;^{3}S_1$&4.167\\
$\eta(1475)$&1475&$\sin\theta_2(\frac{u\bar{u}+d\bar{d}}{\sqrt{2}})+\cos\theta_2({s\bar{s}}),\,\theta_2=?$&$0^{-+}$&$2\;^{1}S_0$&3.301\\
$a_0(1450)$ \cite{Amsler:2008zzb}&1474&$a_0^+=u\bar{d},\,a_0^-=\bar{u}d,\,a_0^0=\frac{u\bar{u}-d\bar{d}}{\sqrt{2}}$&$0^{++}$&$2\;^{3}P_0$&4.347\\
$f^\prime_2(1525)$ \cite{Li:2000zb,Geng:2009iw}&1525&${s\bar{s}}$&$2^{++}$&$1\;^{3}P_2$&3.125\\

$a_2(1700)$ \cite{Amsler:2008zzb}&1732&$a_2^+=u\bar{d},\,a_2^-=\bar{u}d,\,a_2^0=\bar{a_2^0}=\frac{u\bar{u}-d\bar{d}}{\sqrt{2}}$&$2^{++}$&$2\;^{3}P_2$&4.347\\
$\eta(1760)$&1756&$\cos\theta_3(\frac{u\bar{u}+d\bar{d}}{\sqrt{2}})-\sin\theta_3({s\bar{s}}),\,\theta_3=?$&$0^{-+}$&$3\;^{1}S_0$&3.869\\

$X(1835)$&1835&$\sin\theta_3(\frac{u\bar{u}+d\bar{d}}{\sqrt{3}})+\cos\theta_3({s\bar{s}}),\,\theta_3=?$&$0^{-+}$&$3\;^{1}S_0$&3.869\\
$X(2120)$&2120&$\cos\theta_4(\frac{u\bar{u}+d\bar{d}}{\sqrt{2}})-\sin\theta_4({s\bar{s}}),\,\theta_4=?$&$0^{-+}$&$4\;^{1}S_0$&$?$\\
        &&$\sin\theta_4(\frac{u\bar{u}+d\bar{d}}{\sqrt{2}})+\cos\theta_4({s\bar{s}})$&&&\\
$X(2370)$&2370&$\cos\theta_5(\frac{u\bar{u}+d\bar{d}}{\sqrt{3}})-\sin\theta_5({s\bar{s}}),\,\theta_5=?$&$0^{-+}$&$5\;^{1}S_0$&4.348\\
&&$\sin\theta_5(\frac{u\bar{u}+d\bar{d}}{\sqrt{2}})+\cos\theta_5({s\bar{s}})$&&&\\\midrule[1pt]

$K$ \cite{Amsler:2008zzb}&493.68(497.61)&$K^+=u\bar{s},\,K^-=\bar{u}s,\,K^0=d\bar{s},\,\bar{K^0}=\bar{d}s$&$0^-$&$1\;^{1}S_0$&2.174\\
$K^{*}$ \cite{Amsler:2008zzb}&892(896)&$K^{*+}=u\bar{s},\,K^{*-}=\bar{u}s,\,K^{*0}=d\bar{s},\,\bar{K^{*0}}=\bar{d}s$&$1^-$&$1\;^{3}S_1$&3.125\\
$K_1(1270)$ \cite{Amsler:2008zzb}&1272&$K^+_1=u\bar{s},\,K^-_1=\bar{u}s,\,K^0_1=d\bar{s},\,\bar{K^0_1}=\bar{d}s$&$1^+$&$\sin\theta|1\;^{3}P_1\rangle+\cos\theta|1\;^{1}P_1\rangle$&3.448\\
$K_1(1400)$ \cite{Amsler:2008zzb}&1403&$K^+_1=u\bar{s},\,K^-_1=\bar{u}s,\,K^0_1=d\bar{s},\,\bar{K^0_1}=\bar{d}s$&$1^+$&$\cos\theta|1\;^{3}P_1\rangle-\sin\theta|1\;^{1}P_1\rangle$&3.448\\
                                 &    &                                                                         &     &     $\theta=-34^\circ$   \cite{Hatanaka:2008xj}                                   &      \\
$K^{*}(1410)$ \cite{Amsler:2008zzb}&1414&$K^{*+}=u\bar{s},\,K^{*-}=\bar{u}s,\,K^{*0}=d\bar{s},\,\bar{K^{*0}}=\bar{d}s$&$1^-$&$2\;^{3}S_1$&3.846\\
$K^{*}_0(1430)$&1425&$K^{*+}_0=u\bar{s},\,K^{*-}_0=\bar{u}s,\,K^{*0}_0=d\bar{s},\,\bar{K^{*0}_0}=\bar{d}s$&$0^+$&$1\;^{3}P_0$&3.448\\
$K^{*}_2(1430)$ \cite{Amsler:2008zzb}&1425.6(1432.4)&$K^{*+}_2=u\bar{s},\,K^{*-}_2=\bar{u}s,\,K^{*0}_2=d\bar{s},\,\bar{K^{*0}_2}=\bar{d}s$&$2^+$&$1\;^{3}P_2$&3.448\\
$K(1460)$ \cite{Amsler:2008zzb}&1460&$K^+=u\bar{s},\,K^-=\bar{u}s,\,K^0=d\bar{s},\,\bar{K^0}=\bar{d}s$&$0^-$&$2\;^{1}S_0$&3.448\\
$K^{*}(1680)$ \cite{Amsler:2008zzb}&1717&$K^{*+}=u\bar{s},\,K^{*-}=\bar{u}s,\,K^{*0}=d\bar{s},\,\bar{K^{*0}}=\bar{d}s$&$1^-$&$3\;^{3}S_1$&3.846\\

\bottomrule[1pt]
\end{tabular}\caption{The input parameters for the strong decays of $\eta(1295)/\eta(1440)$,
$\eta(1760)/X(1835)$ and $X(2120)/X(2370)$. Other common parameters,
quark masses $m_u=m_d=220$ MeV and $m_s=419$ MeV, are fixed. The
strengths of $q\bar{q}$ and $s\bar{s}$ created from vacuum are
$\gamma_q=6.3$ and $\gamma_s={\gamma_q}/{\sqrt{3}}$, respectively.
The parameter $R$ in the HO wave function is fitted by requiring
reproduction of the realistic root mean square (RMS) radius which is
obtained by solving the Schr\"{o}dinger equation with the potential
in Ref. \cite{Close:2005se} \label{parameter}}
\end{table*}

\begin{thebibliography}{99}

\bibitem{Collaboration:2010au}
  M.~Ablikim {\it et al.}  [BESIII Collaboration],
  arXiv:1012.3510 [hep-ex].

\bibitem{Ablikim:2005um}
  M.~Ablikim {\it et al.}  [BES Collaboration],
  Phys.\ Rev.\ Lett.\  {\bf 95}, 262001 (2005)
  [arXiv:hep-ex/0508025].

\bibitem{Kochelev:2005vd}
  N.~Kochelev and D.~P.~Min,
  Phys.\ Lett.\  B {\bf 633}, 283 (2006)
  [arXiv:hep-ph/0508288].

\bibitem{Kochelev:2005tu}
  N.~Kochelev and D.~P.~Min,
  Phys.\ Rev.\  D {\bf 72}, 097502 (2005)
  [arXiv:hep-ph/0510016].

\bibitem{He:2005nm}
  X.~G.~He, X.~Q.~Li, X.~Liu and J.~P.~Ma,
  Eur.\ Phys.\ J.\  C {\bf 49}, 731 (2007)
  [arXiv:hep-ph/0509140].

\bibitem{Li:2005vd}
  B.~A.~Li,
  Phys.\ Rev.\  D {\bf 74}, 034019 (2006)
  [arXiv:hep-ph/0510093].

\bibitem{Hao:2005hu}
  G.~Hao, C.~F.~Qiao and A.~L.~Zhang,
  Phys.\ Lett.\  B {\bf 642}, 53 (2006)
  [arXiv:hep-ph/0512214].

\bibitem{Bai:2003sw}
  J.~Z.~Bai {\it et al.}  [BES Collaboration],
  Phys.\ Rev.\ Lett.\  {\bf 91}, 022001 (2003)
  [arXiv:hep-ex/0303006].

\bibitem{Datta:2003iy}
  A.~Datta and P.~J.~O'Donnell,
  Phys.\ Lett.\  B {\bf 567}, 273 (2003)
  [arXiv:hep-ph/0306097].

\bibitem{Gao:2003ka}
  C.~S.~Gao and S.~L.~Zhu,
  Commun.\ Theor.\ Phys.\  {\bf 42}, 844 (2004)
  [arXiv:hep-ph/0308205].

\bibitem{Yan:2004xs}
  M.~L.~Yan, S.~Li, B.~Wu and B.~Q.~Ma,
  arXiv:hep-ph/0405087.

\bibitem{Liu:2004er}
  X.~Liu, X.~Q.~Zeng, Y.~B.~Ding, X.~Q.~Li, H.~Shen and P.~N.~Shen,
  arXiv:hep-ph/0406118.

\bibitem{Collaboration:2010zd}
  M.~Ablikim {\it et al.}  [BESIII Collaboration],
  arXiv:1001.5328 [hep-ex].

\bibitem{Alexander:2010vd}
  J.~P.~Alexander {\it et al.}  [CLEO Collaboration],
  Phys.\ Rev.\  D {\bf 82}, 092002 (2010)
  [arXiv:1007.2886 [hep-ex]].

\bibitem{Ding:2005gh}
  G.~J.~Ding and M.~L.~Yan,
  Eur.\ Phys.\ J.\  A {\bf 28}, 351 (2006)
  [arXiv:hep-ph/0511186].

\bibitem{Yan:2006ht}
  M.~L.~Yan,
  arXiv:hep-ph/0605303.

\bibitem{Ding:2007xi}
  G.~J.~Ding and M.~L.~Yan,
  Phys.\ Rev.\  C {\bf 75}, 034004 (2007)
  [arXiv:nucl-th/0702037].

\bibitem{Wang:2006sna}
  Z.~G.~Wang and S.~L.~Wan,
  J.\ Phys.\ G {\bf 34}, 505 (2007)
  [arXiv:hep-ph/0601105].

\bibitem{Liu:2007tj}
  C.~Liu,
  Eur.\ Phys.\ J.\  C {\bf 53}, 413 (2008)
  [arXiv:0710.4185 [hep-ph]].

\bibitem{Dedonder:2009bk}
  J.~P.~Dedonder, B.~Loiseau, B.~El-Bennich and S.~Wycech,
  Phys.\ Rev.\  C {\bf 80}, 045207 (2009)
  [arXiv:0904.2163 [nucl-th]].

\bibitem{Ma:2008hc}
  Y.~L.~Ma,
  J.\ Phys.\ G {\bf 36}, 055004 (2009)
  [arXiv:0808.3764 [hep-ph]].

\bibitem{Huang:2005bc}
  T.~Huang and S.~L.~Zhu,
  Phys.\ Rev.\  D {\bf 73}, 014023 (2006)
  [arXiv:hep-ph/0511153].

\bibitem{Li:2008mza}
  D.~M.~Li and B.~Ma,
  Phys.\ Rev.\  D {\bf 77}, 074004 (2008)
  [arXiv:0801.4821 [hep-ph]].

\bibitem{Amsler:2008zzb}
  C.~Amsler {\it et al.}  [Particle Data Group],
  Phys.\ Lett.\  B {\bf 667}, 1 (2008).

\bibitem{Klempt:2007cp}
  E.~Klempt, A.~Zaitsev,
  Phys.\ Rept.\  {\bf 454}, 1-202 (2007).
  [arXiv:0708.4016 [hep-ph]].

\bibitem{Masoni:2006rz}
  A.~Masoni, C.~Cicalo and G.~L.~Usai,
  J.\ Phys.\ G {\bf 32}, R293 (2006).

\bibitem{Bugg:2011tr}
  D.~V.~Bugg,
  arXiv:1101.1642 [hep-ph].


\bibitem{Gerasimov:2007sb}
  S.~B.~Gerasimov, M.~Majewski and V.~A.~Meshcheryakov,
  J.\ Phys.\ G {\bf 38}, 035008 (2011)
  [arXiv:0708.3762 [hep-ph]].

\bibitem{Cheng:2008ss}
  H.~Y.~Cheng, H.~n.~Li and K.~F.~Liu,
  Phys.\ Rev.\  D {\bf 79}, 014024 (2009)
  [arXiv:0811.2577 [hep-ph]].

\bibitem{Gutsche:2009jh}
  T.~Gutsche, V.~E.~Lyubovitskij and M.~C.~Tichy,
  Phys.\ Rev.\  D {\bf 80}, 014014 (2009)
  [arXiv:0904.3414 [hep-ph]].






\bibitem{Close:1996yc}
  F.~E.~Close, G.~R.~Farrar and Z.~p.~Li,
  Phys.\ Rev.\  D {\bf 55}, 5749 (1997)
  [arXiv:hep-ph/9610280].

\bibitem{Li:2003cn}
  D.~M.~Li, H.~Yu and S.~S.~Fang,
  Eur.\ Phys.\ J.\  C {\bf 28}, 335 (2003).

\bibitem{Li:2007ky}
  G.~Li, Q.~Zhao, C.~-H.~Chang,
  J.\ Phys.\ G {\bf G35}, 055002 (2008).
  [hep-ph/0701020].

\bibitem{Faddeev:2003aw}
  L.~Faddeev, A.~J.~Niemi and U.~Wiedner,
  Phys.\ Rev.\  D {\bf 70}, 114033 (2004)
  [arXiv:hep-ph/0308240].

\bibitem{Ablikim:2006ca}
  M.~Ablikim {\it et al.} [ Bes Collaboration ],
  Phys.\ Rev.\  {\bf D73}, 112007 (2006).
  [hep-ex/0604045].

\bibitem{Bisello:1988as}
  D.~Bisello {\it et al.} [ DM2 Collaboration ],
  Phys.\ Rev.\  {\bf D39}, 701 (1989).

\bibitem{Chew:1962eu}
  G.~F.~Chew and S.~C.~Frautschi,
  Phys.\ Rev.\ Lett.\  {\bf 8}, 41 (1962).

\bibitem{Anisovich:2000kxa}
  A.~V.~Anisovich, V.~V.~Anisovich and A.~V.~Sarantsev,
  Phys.\ Rev.\  D {\bf 62}, 051502 (2000)
  [arXiv:hep-ph/0003113].



\bibitem{Liu:2010tr}
  J.~F.~Liu, G.~J.~Ding and M.~L.~Yan,
  Phys.\ Rev.\  D {\bf 82}, 074026 (2010)
  [arXiv:1008.0246 [hep-ph]].


\bibitem{Micu:1968mk}
  L.~Micu,
  Nucl.\ Phys.\  B {\bf 10}, 521 (1969).

\bibitem{yaouanc}A. Le Yaouanc, L. Oliver, O. P\`{e}ne and J. Raynal,
Phys. Rev. D {\bf 8}, 2223 (1973); D {\bf 9}, 1415 (1974); D {\bf
11}, 1272 (1975); Phys. lett. B {\bf 71}, 57 (1977); B {\bf 71},
397 (1977); .

\bibitem{LeYaouanc:1977gm}
  A.~Le Yaouanc, L.~Oliver, O.~Pene and J.~C.~Raynal,
  Phys.\ Lett.\  B {\bf 72}, 57 (1977).

\bibitem{LeYaouanc:1988fx}
  A.~Le Yaouanc, L.~Oliver, O.~Pene and J.~C.~Raynal,
  ``HADRON TRANSITIONS IN THE QUARK MODEL,''
{\it  NEW YORK, USA: GORDON AND BREACH (1988) 311p}

\bibitem{vanBeveren:1979bd}
  E.~van Beveren, C.~Dullemond and G.~Rupp,
  Phys.\ Rev.\  D {\bf 21}, 772 (1980)
  [Erratum-ibid.\  D {\bf 22}, 787 (1980)];
  E.~van Beveren, G.~Rupp, T.~A.~Rijken and C.~Dullemond,
  Phys.\ Rev.\  D {\bf 27}, 1527 (1983).

\bibitem{Bonnaz:2001aj}
  R.~Bonnaz, B.~Silvestre-Brac and C.~Gignoux,
  Eur.\ Phys.\ J.\  A {\bf 13}, 363 (2002)
  [arXiv:hep-ph/0101112].

\bibitem{sb}W. Roberts and B. Silvestre-Brac, Few-Body Systems, {\bf 11}, 171 (1992).

\bibitem{Blundell:1995ev}
  H.~G.~Blundell and S.~Godfrey,
  Phys.\ Rev.\  D {\bf 53}, 3700 (1996)
  [arXiv:hep-ph/9508264].

\bibitem{Page:1995rh}
  P.~R.~Page,
  Nucl.\ Phys.\  B {\bf 446}, 189 (1995)
  [arXiv:hep-ph/9502204].

\bibitem{Capstick:1986bm}
  S.~Capstick and N.~Isgur,
  Phys.\ Rev.\  D {\bf 34}, 2809 (1986).

\bibitem{Capstick:1993kb}
  S.~Capstick and W.~Roberts,
  Phys.\ Rev.\  D {\bf 49}, 4570 (1994)
  [arXiv:nucl-th/9310030].

\bibitem{Ackleh:1996yt}
  E.~S.~Ackleh, T.~Barnes and E.~S.~Swanson,
  Phys.\ Rev.\  D {\bf 54}, 6811 (1996)
  [arXiv:hep-ph/9604355].

\bibitem{Zhou:2004mw}
  H.~Q.~Zhou, R.~G.~Ping and B.~S.~Zou,
  Phys.\ Lett.\  B {\bf 611}, 123 (2005)
  [arXiv:hep-ph/0412221].

\bibitem{Guo:2005cs}
  X.~H.~Guo, H.~W.~Ke, X.~Q.~Li, X.~Liu and S.~M.~Zhao,
  Commun.\ Theor.\ Phys.\  {\bf 48}, 509 (2007)
  [arXiv:hep-ph/0510146].

\bibitem{Close:2005se}
  F.~E.~Close and E.~S.~Swanson,
  Phys.\ Rev.\  D {\bf 72}, 094004 (2005).

\bibitem{Lu:2006ry}
  J.~Lu, X.~L.~Chen, W.~Z.~Deng and S.~L.~Zhu,
  Phys.\ Rev.\  D {\bf 73}, 054012 (2006)
  [arXiv:hep-ph/0602167].

\bibitem{Zhang:2006yj}
  B.~Zhang, X.~Liu, W.~Z.~Deng and S.~L.~Zhu,
  Eur.\ Phys.\ J.\  C {\bf 50}, 617 (2007)
  [arXiv:hep-ph/0609013].

\bibitem{Chen:2007xf}
  C.~Chen, X.~L.~Chen, X.~Liu, W.~Z.~Deng and S.~L.~Zhu,
  Phys.\ Rev.\  D {\bf 75}, 094017 (2007)
  [arXiv:0704.0075 [hep-ph]];
X. Liu, C. Chen, W.Z.
Deng and X.L. Chen, Chin. Phys. C {\bf 32}, 424-427 (2008) [arXiv:0710.0187 [hep-ph]].

\bibitem{Sun:2009tg}
  Z.~F.~Sun and X.~Liu,
  Phys.\ Rev.\  D {\bf 80}, 074037 (2009)
  [arXiv:0909.1658 [hep-ph]].

\bibitem{Liu:2009fe}
  X.~Liu, Z.~G.~Luo and Z.~F.~Sun,
  Phys.\ Rev.\ Lett.\  {\bf 104}, 122001 (2010)
  [arXiv:0911.3694 [hep-ph]].

\bibitem{Sun:2010pg}
  Z.~F.~Sun, J.~S.~Yu, X.~Liu and T.~Matsuki,
  Phys.\ Rev.\  D {\bf 82}, 111501 (2010)
  [arXiv:1008.3120 [hep-ph]].

\bibitem{Li:2008mzb}
D.~M.~Li and B.~Ma,
  Phys.\ Rev.\  D {\bf 77}, 094021 (2008);
D.~M.~Li and S.~Zhou,
  Phys.\ Rev.\  D {\bf 78}, 054013 (2008);
D.~M.~Li and S.~Zhou,
  arXiv:0811.0918 [hep-ph].

\bibitem{Luo:2009wu}
  Z.~G.~Luo, X.~L.~Chen and X.~Liu,
  Phys.\ Rev.\  D {\bf 79}, 074020 (2009) [arXiv:0901.0505 [hep-ph]].

\bibitem{Jacob:1959at}
  M.~Jacob and G.~C.~Wick,
  Annals Phys.\  {\bf 7} (1959) 404
  [Annals Phys.\  {\bf 281} (2000) 774].

\bibitem{Feldmann:1998vh}
  T.~Feldmann, P.~Kroll and B.~Stech,
  Phys.\ Rev.\  D {\bf 58}, 114006 (1998)
  [arXiv:hep-ph/9802409].

\bibitem{Cheng:2002ai}
  H.~Y.~Cheng,
  Phys.\ Rev.\  D {\bf 67}, 034024 (2003)
  [arXiv:hep-ph/0212117].

\bibitem{Anisovich:2001zp}
  A.~V.~Anisovich, V.~V.~Anisovich and V.~A.~Nikonov,
  Eur.\ Phys.\ J.\  A {\bf 12}, 103 (2001)
  [arXiv:hep-ph/0108186].

\bibitem{Kirchbach:1995ep}
  M.~Kirchbach and D.~O.~Riska,
  Nucl.\ Phys.\  A {\bf 594}, 419 (1995)
  [arXiv:nucl-th/9502018].



\bibitem{Geng:2009iw}
  L.~S.~Geng, F.~K.~Guo, C.~Hanhart, R.~Molina, E.~Oset and B.~S.~Zou,
  Eur.\ Phys.\ J.\  A {\bf 44}, 305 (2010)
  [arXiv:0910.5192 [hep-ph]].

\bibitem{Li:2005eq}
  D.~M.~Li, B.~Ma and H.~Yu,
  Eur.\ Phys.\ J.\  A {\bf 26}, 141 (2005)
  [arXiv:hep-ph/0509215].

\bibitem{Kirchbach:1995ep}
  M.~Kirchbach and D.~O.~Riska,
  Nucl.\ Phys.\  A {\bf 594}, 419 (1995)
  [arXiv:nucl-th/9502018].

\bibitem{Li:2000zb}
  D.~M.~Li, H.~Yu and Q.~X.~Shen,
  J.\ Phys.\ G {\bf 27}, 807 (2001)
  [arXiv:hep-ph/0010342].

\bibitem{Geng:2009iw}
  L.~S.~Geng, F.~K.~Guo, C.~Hanhart, R.~Molina, E.~Oset and B.~S.~Zou,
  Eur.\ Phys.\ J.\  A {\bf 44}, 305 (2010)
  [arXiv:0910.5192 [hep-ph]].

\bibitem{Li:2005eq}
  D.~M.~Li, B.~Ma and H.~Yu,
  Eur.\ Phys.\ J.\  A {\bf 26}, 141 (2005)
  [arXiv:hep-ph/0509215].

\bibitem{Hatanaka:2008xj}
  H.~Hatanaka and K.~C.~Yang,
  Phys.\ Rev.\  D {\bf 77}, 094023 (2008)
  [Erratum-ibid.\  D {\bf 78}, 059902 (2008)]
  [arXiv:0804.3198 [hep-ph]].



\bibitem{Bai:1990hs}
  Z.~Bai {\it et al.}  [MARK-III Collaboration],
  Phys.\ Rev.\ Lett.\  {\bf 65}, 2507 (1990).




\end{thebibliography}
\end{document}